\documentclass[twocolumn]{aastex631}

\newcommand{\degree}{\hbox{$^\circ$}}

\newcommand{\vel}{km\,s$^{-1}$}

\newcommand{\msun}{$M_{\odot}$}
\newcommand{\lsun}{$L_{\odot}$}
\newcommand{\um}{$\mu$m}

\newcommand{\egcite}{\citep[e.g.,][]}

\newcommand{\htcop}{H$^{13}$CO$^{+}$}

\usepackage{ae,aecompl}
\usepackage{subfigure}
\usepackage{graphicx}	
\usepackage{amsmath}	
\usepackage{amssymb}	
\usepackage{color,xcolor}
\usepackage{float}

\received{xxx}
\revised{xxx}
\accepted{xxx}

\shorttitle{Fibers in the IRS\,17 filament}
\shortauthors{Yang et al.}

\begin{document}
\title{\bf The ALMA-QUARKS Survey: Fibers' role in star formation unveiled in an intermediate-mass protocluster region of the Vela D cloud}

\correspondingauthor{Hong-Li Liu}
\email{hongliliu2012@gmail.com}

\author{Dongting Yang}
\affiliation{School of physics and astronomy, Yunnan University, Kunming, 650091, PR China}

\author{Hong-Li Liu}
\affiliation{School of physics and astronomy, Yunnan University, Kunming, 650091, PR China}

\author{Tie Liu}
\affiliation{Shanghai Astronomical Observatory, Chinese Academy of Sciences, 80 Nandan Road, Shanghai 200030, Peoples Republic of China}
\correspondingauthor{Tie Liu}
\email{liutie@shao.ac.cn}

\author{Anandmayee Tej}
\affiliation{Indian Institute of Space Science and Technology, Thiruvananthapuram 695 547, Kerala, India}
\correspondingauthor{Anandmayee Tej}
\email{tej@iist.ac.in}

\author{Xunchuan Liu}
\affiliation{Shanghai Astronomical Observatory, Chinese Academy of Sciences, 80 Nandan Road, Shanghai 200030, Peoples Republic of China}
\correspondingauthor{Xunchuan Liu}
\email{liuxunchuan001@gmail.com}

\author{Jinhua He}
\affiliation{Yunnan Observatories, Chinese Academy of Sciences, Kunming, 650216, Yunnan, PR China}

\author{Guido Garay}
\affiliation{Departamento de Astronom\'ia, Universidad de Chile, Casilla 36-D, Santiago, Chile}
\affiliation{Chinese Academy of Sciences South America Center for Astronomy, National Astronomical Observatories, CAS, Beijing 100101, China}

\author{Amelia Stutz}
\affiliation{Departamento de Astronom\'ia, Universidad de Concepci\'on, Av. Esteban Iturra s/n, Distrito Universitario, 160-C, Chile}
\affiliation{Max Planck Institute for Astronomy, Königstuhl 17, D-69117 Heidelberg, Germany}

\author{Lei Zhu}
\affiliation{Chinese Academy of Sciences South America Center for Astronomy, National Astronomical Observatories, Chinese Academy of Sciences, Beijing, 100101, PR China}

\author{Sheng-Li Qin}
\affiliation{School of physics and astronomy, Yunnan University, Kunming, 650091, PR China}

\author{Fengwei Xu}
\affiliation{Kavli Institute for Astronomy and Astrophysics, Peking University, 5 Yiheyuan Road, Haidian District, Beijing 100871, People's Republic of China}
\affiliation{Department of Astronomy, Peking University, 100871, Beijing, People's Republic of China}

\author[0000-0001-8077-7095]{Pak-Shing Li}
\affiliation{Shanghai Astronomical Observatory, Chinese Academy of Sciences, 80 Nandan Road, Shanghai 200030, Peoples Republic of China}

\author[0000-0002-5809-4834]{Mika Juvela}
\affiliation{Department of Physics, P.O. box 64, FI- 00014, University of Helsinki, Finland}

\author{Pablo Garc\'ia}
\affiliation{Chinese Academy of Sciences South America Center for Astronomy, National Astronomical Observatories, CAS, Beijing 100101, China}
\affiliation{Instituto de Astronom\'ia, Universidad Cat\'olica del Norte, Av. Angamos 0610, Antofagasta, Chile}

\author{Paul F. Goldsmith}
\affiliation{Jet Propulsion Laboratory, California Institute of Technology, 4800 Oak Grove Drive, Pasadena, CA 91109, USA}

\author{Siju Zhang}
\affiliation{Departamento de Astronom\'{i}a, Universidad de Chile, Las Condes, 7591245 Santiago, Chile}

\author{Xindi Tang}
\affiliation{Xinjiang Astronomical Observatory, Chinese Academy of Sciences,
150 Science 1-Stree, Urumqi, Xinjiang 830011, People's Republic of China}

\author{Patricio Sanhueza}
\affiliation{Department of Earth and Planetary Sciences, Tokyo Institute of Technology, Meguro, Tokyo, 152-8551, Japan}
\affiliation{National Astronomical Observatory of Japan, National Institutes of Natural Sciences, 2-21-1 Osawa, Mitaka, Tokyo 181-8588, Japan}

\author[0000-0003-1275-5251]{Shanghuo Li}
\affiliation{Max Planck Institute for Astronomy, Königstuhl 17, D-69117 Heidelberg, Germany}

\author{Chang Won Lee}
\affiliation{Korea Astronomy and Space Science Institute, 776 Daedeokdae-ro, Yuseong-gu, Daejeon 34055, Republic of Korea}
\affiliation{University of Science and Technology, Korea (UST), 217 Gajeong-ro, Yuseong-gu, Daejeon 34113, Republic of Korea}

\author{Swagat Ranjan Das}
\affiliation{Departamento de Astronom\'ia, Universidad de Chile, Casilla 36-D, Santiago, Chile}

\author{Wenyu Jiao}
\affiliation{Shanghai Astronomical Observatory, Chinese Academy of Sciences, 80 Nandan Road, Shanghai 200030, Peoples Republic of China}

\author[0000-0002-0786-7307]{Xiaofeng Mai}
\affiliation{Shanghai Astronomical Observatory, Chinese Academy 
of Sciences, Shanghai 200030, People’s Republic of China}
\affiliation{School of Astronomy and Space Sciences, University of
Chinese Academy of Sciences, No. 19A Yuquan Road, Beijing 100049,
People’s Republic of China}

\author{Prasanta Gorai}
\affiliation{Rosseland Centre for Solar Physics, University of Oslo, PO Box 1029 Blindern, 0315 Oslo, Norway}
\affiliation{Institute of Theoretical Astrophysics, University of Oslo, PO Box 1029 Blindern, 0315 Oslo, Norway}

\author{Yichen Zhang}
\affiliation{Department of Astronomy, Shanghai Jiao Tong University, 800 Dongchuan Rd., Minhang, Shanghai 200240, People's Republic of China }

\author{Zhiyuan Ren}
\affiliation{National Astronomical Observatories, Chinese Academy of Sciences, Datun Road A20, Beijing, People's Republic of China}
\affiliation{CAS Key Laboratory of FAST, NAOC, Chinese Academy of Sciences, Beijing, China}
\affiliation{University of Chinese Academy of Sciences, Beijing, People's Republic of China}

\author[0000-0002-5310-4212]{L. Viktor T\'oth}
\affiliation{Institute of Physics and Astronomy, E\"otv\"os Lor\`and University, P\'azm\'any P\'eter s\'et\'any 1/A, H-1117 Budapest, Hungary}

\author{Jihye Hwang}
\affiliation{Korea Astronomy and Space Science Institute, 776 Daedeokdae-ro, Yuseong-gu, Daejeon 34055, Republic of Korea}

\author{Leonardo Bronfman}
\affiliation{Departamento de Astronom\'ia, Universidad de Chile, Casilla 36-D, Santiago, Chile}

\author{Ken'ichi Tatematsu}
\affiliation{National Astronomical Observatory of Japan, National Institutes of Natural Sciences, 2-21-1 Osawa, Mitaka, Tokyo 181-8588, Japan}

\author{Lokesh Dewangan}
\affiliation{Physical Research Laboratory, Navrangpura, Ahmedabad—380 009, India}

\author[0000-0002-9875-7436]{James O. Chibueze}
\affiliation{Department of Mathematical Sciences, University of South Africa, Cnr Christian de Wet Rd and Pioneer Avenue, Florida Park, 1709, Roodepoort, South Africa}
\affiliation{Department of Physics and Astronomy, Faculty of Physical Sciences, University of Nigeria, Carver Building, 1 University Road, Nsukka 410001, Nigeria}

\author{Suinan Zhang}
\affiliation{Shanghai Astronomical Observatory, Chinese Academy of Sciences, 80 Nandan Road, Shanghai 200030, Peoples Republic of China}

\author{Gang Wu}
\affiliation{Xinjiang Astronomical Observatory, Chinese Academy of Sciences,
150 Science 1-Stree, Urumqi, Xinjiang 830011, People's Republic of China}

\author{Jinjin Xie}
\affiliation{Shanghai Astronomical Observatory, Chinese Academy of Sciences, 80 Nandan Road, Shanghai 200030, Peoples Republic of China}

\begin{abstract}
In this paper, we present a detailed analysis of the IRS\,17 filament within the intermediate-mass protocluster IRAS 08448-4343 (of $\sim\,10^3$\,\lsun), using ALMA data from the ATOMS 3-mm and QUARKS 1.3-mm surveys. The IRS\,17 filament, which spans $\sim$54000\,au ($0.26\,\rm pc$) in length and $\sim$4000\,au ($0.02\,\rm pc$) in width, exhibits a complex, multi-component velocity field, and harbours hierarchical substructures. These substructures include three bundles of seven velocity-coherent fibers, and 29 dense ($n\sim 10^8\,\rm cm^{-3}$) condensations.
The fibers have a median length of $\sim 4500\,\rm au$ and a median width of $\sim 1400\,\rm au$. Among these fibers, four are identified as ``fertile", each hosting at least three dense condensations, which are regarded as the ``seeds" of star formation. While the detected cores are randomly spaced within the IRS\,17 filament based on the 3-mm dust continuum image, periodic spacing ($\sim1600\,\rm au$) of condensations is observed in the fertile fibers according to the 1.3-mm dust map, consistent with the predictions of linear isothermal cylinder fragmentation models.
These findings underscore the crucial role of fibers in star formation and suggest a hierarchical fragmentation process that extends from the filament to the fibers, and ultimately, to the smallest-scale condensations.

\end{abstract}


\section{Introduction} \label{sec:intro}
Filamentary structuresare prevalent in the molecular clouds of the Milky Way, and recognized to play a vital role in both low-mass and high-mass star formation scenarios \citep{2010A&A...518L.100M,2010A&A...518L.102A,2014prpl.conf...27A,2020A&A...635A..34K,2023MNRAS.522.3719L}. {\it Herschel} dust continuum observations in far/submillimeter bands have demonstrated that prestellar cores and protostars predominantly form in the densest filaments within molecular clouds \egcite{2010A&A...518L.106K,2016A&A...590A...2S,2017A&A...600A.141K,2018MNRAS.478.2119L,2019MNRAS.487.1259L}.
Moreover, filaments resolved in nearby star-forming regions exhibit a power-law distribution for both the filament mass function (FMF), \mbox{$\frac{\mathrm{d}N}{\mathrm{d}M} \propto M^{-2.6}$}, and filament line mass function (FLMF), \mbox{$\frac{\mathrm{d}N}{\mathrm{d}m} \propto m^{-2.4}$}, respectively \citep{2019A&A...629L...4A}. This may naturally result in a power-law distribution of core masses \citep{2019A&A...629L...4A,2019A&A...621A..42A}, i.e., the core mass function (CMF). Accordingly, the CMF has also been observed to agree well with the stellar initial mass function (IMF) in the range of low to intermediate stellar masses ($\sim 0.1-5$\,\msun, \citealt{2010A&A...518L.106K}), as previously reported \egcite{1998A&A...336..150M,2007ASSP....3..417A,2007A&A...472..519A}.
Nonetheless, this agreement has been observed to be challenging in regions of massive star formation \egcite{2015ApJ...804..141Z,2018ARA&A..56...41M,2019ApJ...886..102S,2022A&A...664A..26P,2023A&A...674A..76P,2024ApJ...961L..35M,2024arXiv240718719L}, with an evident deviation between the CMF and the IMF in terms of the slope at the high-mass end. In this context, filaments play a critical role in the star formation process, setting up its initial conditions.
However, their formation, evolution into substructures, dense cores, and ultimately stars, is still not fully understood.

\begin{figure}[ht!]
    \centering
    \includegraphics[angle=0, width=0.45\textwidth]{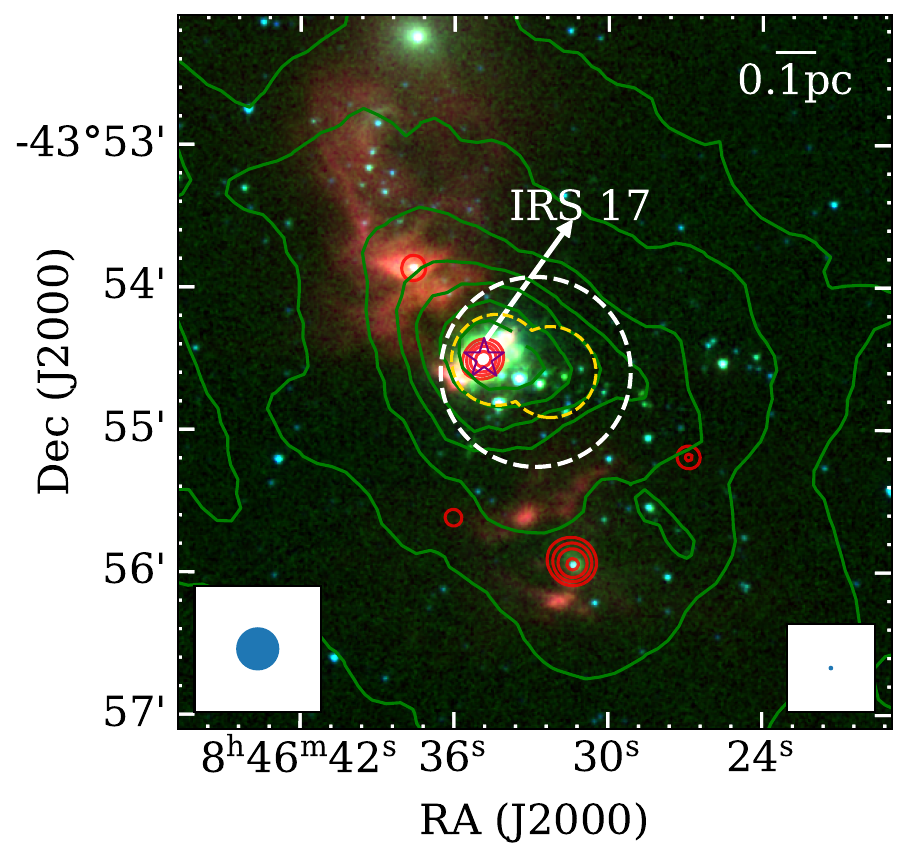} 
    \caption{{\it Spitzer} three-colour composite image using 
    8\,\um\,(red), 4.5\,\um\,(green) and 3.6\,\um\,(blue) of the IRS\,17 region. The {\it Herschel} 250\,\um\, and 1.28\,GHz MeerKAT emission
    \citep{2023MNRAS.524.1291P} are shown as the green and red contours, respectively. The white and yellow dashed circle represent the field of views (FoV) of the ATOMS and QUARKS data, respectively (see Fig.\,\ref{fig:band3:con} ). The purple star symbol shows the location of IRS\,17 source.  The resolutions of 8\,\um\, and 250\,\um\, are shown at the bottom right and bottom left corners, respectively.}
    \label{fig:RGB}    
\end{figure}

Substantial progress has been made in deciphering the evolution of filaments. When scrutinized at a high enough resolution, an increasing number of studies reveal a new type of dense molecular filaments at small scales, which are called dense fibers. These fibers are detected across the mass spectrum of clouds, for example, in low-mass clouds \citep{2013A&A...554A..55H,2013A&A...553A.119A,2016A&A...590A..75F}, intermediate mass clusters \citep{2014ApJ...790L..19F,2017A&A...606A.123H}, and high-mass star-forming regions \citep{2018A&A...610A..77H,2019A&A...629A..81T,2019A&A...623A..16S,2022ApJ...926..165L,2022ApJ...927..106C}. As summarized by \cite{2023ASPC..534..153H}, the fibers typically display sub-parsec lengths ($L\lesssim$ 1pc), and have line masses near the critical value \citep{2018A&A...610A..77H} for hydrostatic equilibrium. Fibers also exhibit high central gas densities \mbox{$n_0 > 10^4$ cm$^{-3}$}, and small characteristic widths (full width half maximum (FWHM)$<0.1$\,pc; see \citealt{2018A&A...610A..77H}, measured in the dense gas tracer, $\rm N_2H^+$\,(1-0). 
It is interesting to note that the width of filaments or fibers depends on the gas tracer investigated. The widths are observed to be narrower for dense gas tracers, such as \htcop $(1-0)$ and $\rm N_2H^+$\,(1-0) as compared to the low-density gas tracers like $\rm ^{13}CO\,(1-0)$ and $\rm C^{18}O\,(1-0)$ \citep{2023A&A...672A.133S}.

Fibers correspond to the hierarchical substructure within larger and more massive filaments \citep[e.g.][]{2014prpl.conf...27A}. 
Whether fibers or larger filaments form first remains a subject of debate due to the existence of two possibilities. That is, hierarchical fiber structures could arise either from top-down fragmentation due to the self-gravity of their parent filament \citep{2013A&A...554A..55H,2015A&A...574A.104T}, or from a bottom-up process whereby gravitational collapse assembles larger filaments from gas containing pre-existing fibers \citep{2014MNRAS.445.2900S}. 

In addition to dense fibers, dense cores are viewed as crucial substructures of filaments where stars form.  
The spacing of prestellar cores in star-forming filaments is generally observed to be non-periodic \citep[e.g.,][]{2014prpl.conf...27A,2020A&A...635A..34K}, though a few examples of quasi-periodic chains of dense cores have been reported \citep{2015A&A...574A.104T,2020A&A...642A..76Z, 2019MNRAS.487.1259L,2023ApJ...951...68C,2023ApJ...957...94Y}. Theoretically, periodic core spacing is predicted by standard semi-analytic cylinder fragmentation models within the framework of gravitational instability of filaments (e.g., \citealt{1963AcA....13...30S,2010ApJ...719L.185J,2018MNRAS.478.2119L}).
This discrepancy between observations and simple theoretical predictions may be attributed to the presence of hierarchical fibers within larger filaments \citep{2020MNRAS.497.4390C}. 
Therefore, examining the substructures (i.e., fibers, and cores) of parental filaments is beneficial in understanding the evolution of filaments into cores and ultimately stars.

The target of this work is the IRAS\,08448-4343 protocluster (see Fig.\,\ref{fig:RGB}), which is part of the Vela Molecular Ridge-D (VMR-D) cloud \citep{1991A&A...247..202M} located at $\sim 1.2\pm0.1$\,kpc \citep{2024RAA....24b5009L}. This distance is consistent with
the range of 0.7 to 1.2\,kpc estimated by \citet{1992A&A...265..577L} using photometric distance methods. The IRAS\,08448-4343 protocluster clump has a molecular gas mass of $\sim 116$\,\msun\  within a size of 0.5\,pc, which have been scaled to the distance adopted here (e.g., \citealt{2004A&A...426...97F,2020MNRAS.496.2790L}).
This protocluster corresponds to a bright infrared (IR) source,  IRS\,17, identified by \citet{1992A&A...265..577L} as an intermediate-mass Class\,I YSO object according to its  luminosity of $\sim 10^3$\,\lsun\  \citep{1992A&A...265..577L,1999A&AS..136..471M}. Near-IR observations reveal that more than 20 jet knots are associated with IRS\,17 ($\#\,57$ in \citealt{2002ApJ...564..839L} and \citealt{2005A&A...433..941G}). The {\it Spitzer} three-color image in 3.6/4.5/8.0\,\um\, (see Fig.\,\ref{fig:RGB}) displays an IR-bright patch and point sources surrounding IRS\,17. 
Furthermore, the IRS\,17 region appears as a compact source in radio centimeter emission (e.g., at 6\,cm, \citealt{2007A&A...476.1019M}), which is confirmed by 1.28\,GHz radio emission (see red contours in Fig.\,\ref{fig:RGB}), retrieved from the MeerKAT Galactic Plane Survey \citep{2023MNRAS.524.1291P,2024MNRAS.531..649G}. 
This suggests strong feedback from ionized gas excited by the IRS\,17 protocluster. 
From {\it Herschel} 250\,\um\ emission, the molecular cloud hosting the IRS\,17 region appears as a clumpy structure elongated along the northeast-southwest direction. This elongation is manifested as a filamentary structure
(hereafter the IRS\,17 filament) when observed in
ALMA 3-mm continuum image at a resolution of $\sim$2\,\arcsec\ \citep{2024RAA....24b5009L}. 

In this work, we investigate the role of small-scale fibers in the evolution of the IRS\,17 filament into cores/condensations, leveraging on the combination of the ALMA 3-mm  and our new 1.3-mm observations (see below). This paper is structured as follows:  observations are described in Section\,\ref{sec:obse}; the results and analysis are presented in Section\,\ref{sec:result}; the discussion is followed in Section\,\ref{sec:dsicussion}; and finally the summary and conclusion are given in Section\,\ref{sec:summary}.  

\begin{figure}[ht!]
    \centering
    \includegraphics[angle=0, width=0.48\textwidth]{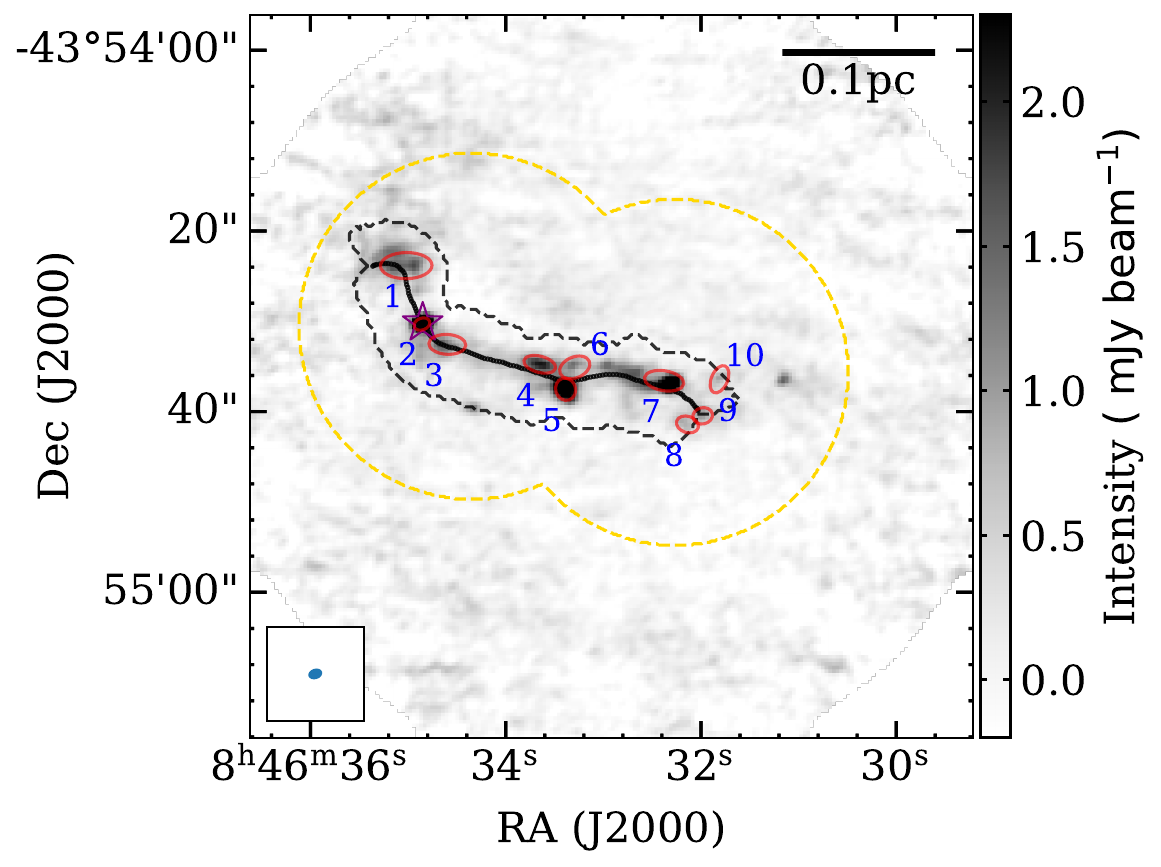} 
    \caption{3-mm dust continuum image of the IRS\,17 protocluster region from the ATOMS survey. The black solid and dotted lines respectively indicate the footprint and skeleton of the IRS\,17 filament obtained with the {\it getsf} algorithm on the 3-mm dust continuum data. The yellow  overlapping circles represent the FoV of the QUARKS survey and the red ellipses represent 10 dense cores within the IRS\,17 filament footprint, which were identified by \cite{2021MNRAS.505.2801L}. }
    \label{fig:band3:con}    
\end{figure}

\begin{figure*}[t]
    \centering
    \includegraphics[angle=0, width=0.9\textwidth]{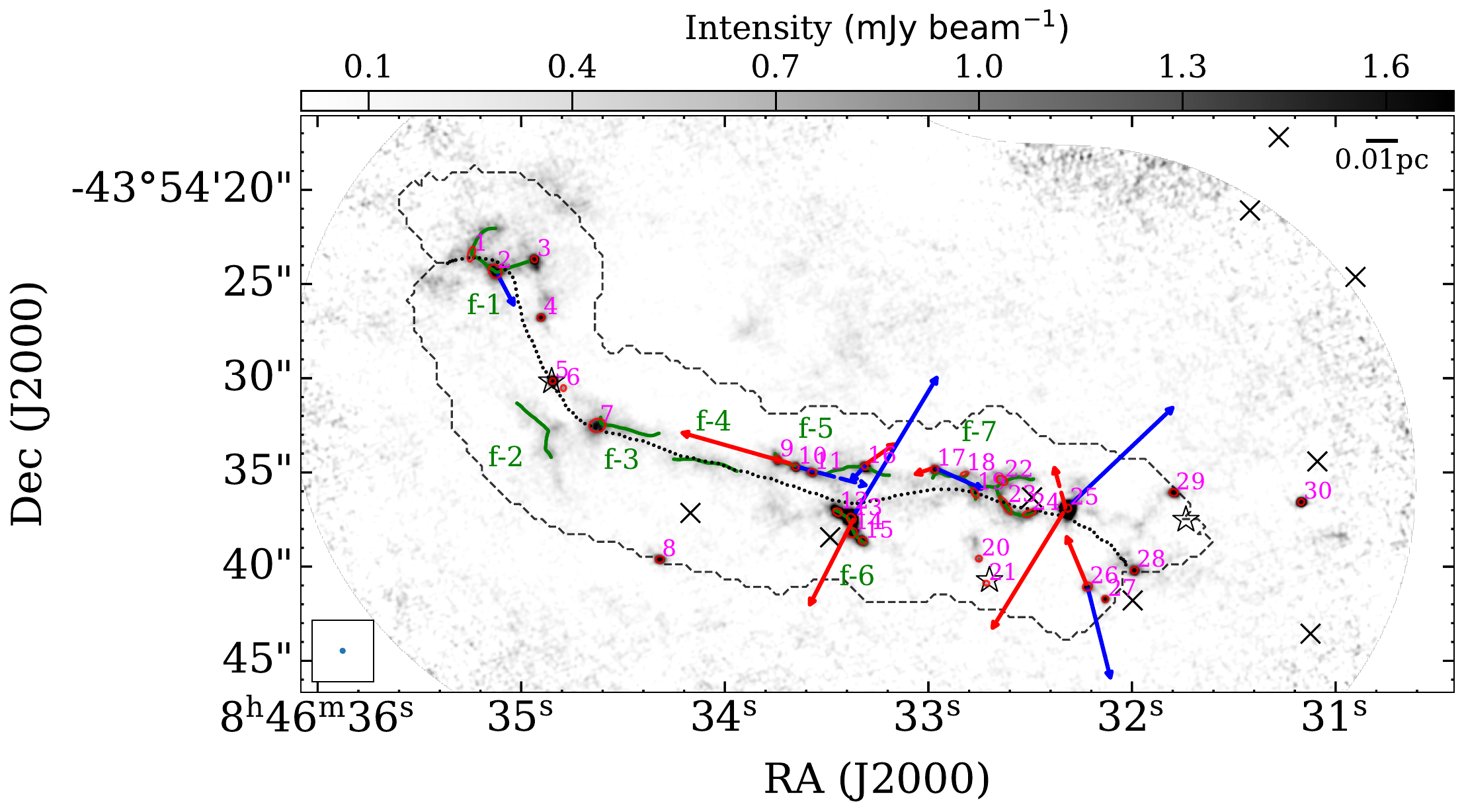} 
    \caption{1.3-mm dust continuum image from the QUARKS survey. The black dotted line shows the skeleton of the IRS\,17 filament, and green solid lines represent the dense fibers. Red ellipses show the condensations identified by the {\it getsf} and SExtractor algorithm. The black stars and crosses are the YSO Class\,I/0 objects and IRAC point-like sources as candidate YSOs \citep{2010ApJ...719....9S,2015ApJ...798..104S}, respectively. The solid and dashed arrows represent the outflows identified either from CO (2--1) or SiO (5--4), respectively (see Fig.\,\ref{fig:outflow}). The synthesized beam size and the scale-bar are shown in left bottom and top right, respectively.}
    \label{fig:bad6}    
\end{figure*}

\section{ALMA observations}\label{sec:obse}
The IRS\,17 region was observed as part of the ALMA Three-millimeter Observations of Massive Star-forming regions (ATOMS) survey, which targeted 146 active massive protoclusters regions with the ALMA 7-m Atacama Compact Array (ACA) and the 12-m array in the 3-mm band \citep{2020MNRAS.496.2790L}. The data were calibrated in CASA 5.6 \citep{2007ASPC..376..127M}. The ACA and 12-m array data were jointly imaged and cleaned using natural weighting for pblimit = 0.2, in the CASA tclean task, for both continuum images and line cubes. The combined data for the IRS\,17 region have a field of view (FoV) and a maximum recoverable scale both of $\sim80\arcsec$ or 0.5\,pc at the distance of IRS\,17 (e.g.; $1.2\pm0.1\,\rm kpc$).
The combined continuum data have a synthesized beam size of $1.6\arcsec \times 1.2\arcsec$, and a typical sensitivity of $rms\sim0.2$\,mJy~beam$^{-1}$. 
The \htcop~(1--0) line cube, used here for kinematics analysis, has a synthesized beam size of $1.8\arcsec \times 1.4\arcsec$ and a sensitivity of $rms \sim8$\,mJy~beam$^{-1}$ at a velocity resolution of $\rm \sim 0.2\,km~s^{-1}$. 
Details on the observing setups of the survey and its data reduction can be found in \cite{2020MNRAS.496.2790L,2020MNRAS.496.2821L,2021MNRAS.505.2801L,2022MNRAS.510.5009L,2022MNRAS.511.4480L}.

IRS\,17 was also observed as part of the Querying Underlying mechanisms of massive star formation with ALMA-Resolved gas Kinematics and Structures (QUARKS) survey, complementing the ATOMS survey with higher angular resolution ($\sim 0.3\arcsec$). The observations were carried out with both the ACA and 12-m arrays \citep{2024RAA....24b5009L}. In particular, the elongated spatial extent of the IRS\,17 region was covered by two pointings, each of which has a field of view of 40\arcsec or 0.2\,pc (see the yellow curve in Fig.\,\ref{fig:band3:con}), and a maximum recoverable scale of $\sim20\,\arcsec$ or 0.1\,pc.  The ACA and 12-m data were independently calibrated in CASA\,6.5, and then jointly imaged and cleaned using the CASA tclean task for both continuum images and line cubes. The combined ACA and 12-m continuum data for the IRS\,17 region  have a synthesized beam size of $0.34\arcsec \times 0.33\arcsec$, corresponding to $\sim$400\,au at a distance of the source. The sensitivity of the continuum data is $\sim0.15\,\rm\,mJy\,beam^{-1}$, which translates to a mass sensitivity of $\sim 10^{-3}$\,\msun\, for a dust temperature of $\sim 25$\,K (see Sect.\,\ref{subsub:sec:fibers and condensation} for the detailed mass calculation). Moreover, the QUARKS survey includes the data of major lines at 1.3\,mm, such as $\rm N_2D^+$, DCN, $\rm H_2CO$, and $\rm ^{13}CS$ (see Table\,1 of \citealt{2024RAA....24b5009L}). 
The native velocity resolution of these line data is 0.6--0.7\,\vel. However, these line data have low detection rates in the IRS\,17 region, and hence are considered only for ancillary analysis.
Details on the observing setups of the survey and data reduction are described in \cite{2024RAA....24b5009L} and \cite{2024arXiv240402275X}

\begin{deluxetable*}{lccccccccc}
\tabletypesize{\footnotesize}
\tablecaption{Parameters of filament and fibers obtained with the {\it getsf} algorithm \label{tab:filament}}
\tablewidth{0pt}
\tablehead{
\colhead{Name} & \multicolumn{2}{c}{\underline {~~~~~~~~~~$\rm Position^{\it a}$~~~~~~~~~~}}     & \multicolumn{2}{c}{\underline {~~~~~~~~~~$ L^{\it b}$~~~~~~~~~~}}  & \multicolumn{2}{c}{\underline {~~~~~~~~~~$ W^{\it b}$~~~~~~~~~~}}  & \colhead{Aspect ratio} & \colhead{$S^{\rm int}_{\nu}$}   & \colhead{Data Source} \\
\colhead{} &  \colhead{$\alpha$(J2000)} & \colhead{$\delta$(J2000)}& \colhead{(\arcsec)}& \colhead{$(\rm \times 10^3\,au)$}&\colhead{(\arcsec)} &\colhead{$(\rm \times 10^3\,au)$} & \colhead{} & \colhead{(mJy)} & \colhead{} }
\startdata
 Filament  &    08:46:33.84  & -43:54:32.53   &  $43.5\pm4.4$  &	$53.6\pm5.4$ 	&  $3.2\pm1.5$  & $3.9\pm1.9$ &  13.6 &  $43.8\pm4.4$     &  3\,mm 	\\
    \hline
    fiber	  & &  &    &     &       &       &  &    &   1.3\,mm       \\
  1         & 08:46:35.12  & -43:54:23.44&  $5.5\pm0.6$  &	$6.7\pm0.7$ 	&   $1.5\pm0.6$  &  $1.9\pm0.8$    &  3.7 &  $103.8\pm10.4$     &  	\\
  2         & 08:46:34.92  & -43:54:32.46&  $3.8\pm0.4$  &	$4.6\pm0.5$ 	&   $1.2\pm0.3$  &  $1.5\pm0.4$    &  3.1 &  $34.8\pm3.5$     &  	\\
  3         & 08:46:34.47  & -43:54:33.00&  $3.5\pm0.4$  &	$4.3\pm0.4$ 	&   $1.2\pm0.5$  &  $1.5\pm0.6$    &  3.0 &  $44.5\pm4.5$     &  	\\
  4         & 08:46:34.05  & -43:54:34.54&  $3.5\pm0.4$  &	$4.3\pm0.4$ 	&   $1.1\pm0.3$  &  $1.3\pm0.4$    &  3.5 &  $27.2\pm2.7$     &  	\\
  5         & 08:46:33.49  & -43:54:34.66&  $6.6\pm0.7$  &	$8.0\pm0.8$ 	&   $0.9\pm0.6$  &  $1.1\pm0.8$    &  7.8 &  $78.9\pm7.9$     &  	\\
  6         & 08:46:33.40  & -43:54:38.20&  $2.6\pm4.4$  &	$3.2\pm0.3$ 	&   $0.8\pm0.2$  &  $1.0\pm0.3$    &  3.2 &  $50.5\pm5.1$     &  	\\
  7         & 08:46:32.69  & -43:54:36.17&  $7.1\pm0.7$  &	$8.7\pm0.9$ 	&   $1.3\pm0.8$  &  $1.5\pm0.9$    &  5.8 &  $111.5\pm11.2$     &  	\\
\enddata
\begin{flushleft}
$^a$ The central coordinate of IRS\,17 filament and fibers.\\
$^b$ The lengths and widths are derived from the footprints of filament and the fibers extracted with the {\it getsf} algorithm. 
\end{flushleft}
\end{deluxetable*}

\begin{deluxetable*}{lcccccc}
\tabletypesize{\footnotesize}
\tablecaption{Estimated properties and velocity information of filament and fibers \label{tab:filament:calculated}}
\tablewidth{0pt}
\tablehead{
\colhead{Name} & \colhead{$M$}     & \colhead{$m$} & \colhead{$m^{\rm tur}_{\rm crit}$} & \colhead{$\langle n \rangle$} &  \colhead{$ v_{\rm lsr}^*$ }  &  \colhead{$\sigma _{\rm obs}^*$ }  \\
\colhead{} &  \colhead{(\msun)} & \colhead{(\msun $\rm ~pc^{-1}$)}& \colhead{(\msun $\rm ~pc^{-1}$)}& \colhead{$\rm (\times 10^7\,cm^{-3})$}& \colhead{$\rm (km\,s^{-1})$}&\colhead{$\rm (km\,s^{-1})$} }
\startdata
  Filament  &    $24.9\pm5.0$  & $95.8\pm19.2$& $70.7\pm29.1$   &  $0.49\pm0.15$  &	$3.26\pm1.15$ &  $0.39\pm0.25$ \\
     \hline
    fiber	 &&&&&&   \\
  1         & $2.54\pm0.51$  &  $77.0\pm15.4$ &  $178.7\pm47.6$ &  $1.70\pm0.34$ &	$5.11\pm0.30$ & $0.62\pm0.32$ 	\\
  2         & $0.83\pm0.17$ &   $37.7\pm7.5$ &   $67.1\pm5.6$  & $1.29\pm0.10$ &	$5.31\pm0.20$ & $0.38\pm0.11$   	\\
  3         & $1.09\pm0.22$ &   $51.9\pm10.4$ &  $50.6\pm29.1$   & $1.82\pm0.36$  &	$3.25\pm0.87$  &  $0.33\pm0.25$	\\
  4         &  $0.66\pm0.13$ &   $31.4\pm6.3$  &  $47.6\pm22.5$ & $1.47\pm0.15$   &	 $3.11\pm0.79$  &  $0.32\pm0.22$    \\
  5         & $1.92\pm0.38$  &   $49.2\pm9.8$ &   $70.7\pm24.6$ & $3.20\pm1.60$   &	$3.39\pm0.27$  & $0.39\pm0.23$  	\\
  6         & $1.23\pm0.25$ &   $82.0\pm16.4$ &   $90.0\pm7.9$  &$6.21\pm0.62$   &	 $3.10\pm0.40$   &  $0.44\pm0.13$    \\
  7         & $2.72\pm0.54$ &  $61.8\pm12.4$  &   $94.2\pm9.1$ & $2.24\pm0.89$   &	$2.33\pm0.64$  &  $0.45\pm0.14$ 	\\
\enddata
\begin{flushleft}
$^*$\,The velocity centroid and velocity dispersion measured from the \htcop\ line emission of the ATOMS survey. 
\end{flushleft}
\end{deluxetable*}

\section{Results and analysis}\label{sec:result}
\subsection{IRS\,17 filament seen in ATOMS 3-mm continuum}
\label{subsec:filament:source}
Figure\,\ref{fig:band3:con} shows the 3-mm continuum map of the IRS\,17 protocluster region (also refer to Fig.\,\ref{fig:band3:band6}a). The IRS\,17 filament extends from the northeast to the southwest. To identify this filament, the {\it getsf} extraction algorithm \citep{2021A&A...649A..89M} was used. This algorithm has been widely employed in previous studies \citep{2023MNRAS.520.3259X,2024arXiv240402275X,2024MNRAS.527.5895D,2024arXiv240718719L} due to its effectiveness in handling uneven noise background, separating blended sources/filaments, and extracting extended emission features. After applying the {\it getsf} to the 3-mm continuum map, we obtained a single filament skeleton and its footprint, that are shown in Fig.\,\ref{fig:band3:con} as the black line and the dashed contour, respectively. 
The filament parameters measured by the {\it getsf} algorithm are listed in Table\,\ref{tab:filament}. These include the center positions, length, width, and integrated flux. 
The IRS\,17 filament has a length of $\sim 5\times10^4\,\rm au$ (0.26\,pc) and a width of  $\sim 4\times 10^3\,\rm au$ (0.02\,pc). The tablulated widths
refer to the mean values of the median widths obtained from both side by the {\it getsf} algorithm. Note that the median width corresponds to half-power diameters estimated directly from the profiles using {\it getsf}, without any fitting, such as Gaussian or Plummer fitting \citep{2021A&A...649A..89M}. 

Assuming that 3-mm continuum emission is optically thin and  arises primarily from thermal dust radiation, we can compute the mass of the IRS\,17 filament using the expression

\begin{equation}
\label{equ:mass:cal}
    M=\frac{S^{\rm int}_\nu D^2R_{\rm gd}}{\kappa_\nu B_\nu(T_{\rm dust})},
\end{equation}
where the integrated flux, $S^{\rm int}_\nu$, is derived from 3-mm continuum emission,   {\it D} is the source distance, the ratio of gas-to-dust, $R_{\rm gd}$, is assumed to be 100, and the dust opacity, $\kappa_\nu$, is taken as $0.18 \rm\,cm^{2}~g^{-1}$ at 3\,mm \citep{1994A&A...291..943O,2021MNRAS.505.2801L}. $B_\nu(T_{\rm dust}$) is the Planck function for a given dust temperature. Here, $T_{\rm dust}= \rm 25\,K$ is adopted for the IRS\,17 filament, the same as its natal clump's average temperature derived from the spectral energy distribution \citep{2004A&A...426...97F}. 

The calculated mass of the IRS\,17 filament is $M_{\rm F}\sim 24.9$\,\msun. The line mass ($m_{\rm F}$) was computed as $m_{\rm F} = M_{\rm F}/L_{\rm F}\sim95.8$\,\msun\,$\rm pc^{-1}$.
This line mass is around 2–3 times higher than those inferred from {\it Herschel} observations toward low mass star forming filaments, such as Musca (see Fig.\,4 of \citealt{2016A&A...590A.110C}), and Taurus B211/3 (see Fig.\,2 of  \citealt{2013A&A...550A..38P}) but lower than those toward high-mass counterparts such as Orion\,ISF \citep{2016A&A...590A...2S}. 
Assuming that the FMF could naturally produce the CMF, which in turn is proportional to the IMF \citep{2019A&A...629L...4A}, this result suggests that the IRS\,17 filament is an intermediate-mass star forming region. Additionally, it is also consistent with the presence of the intermediate-mass IR bright YSO (i.e., the IRS\,17 point source). Moreover, assuming a uniform density distribution for the filament, the average number density ($\langle n_{\rm F} \rangle$, see Col.4 in Table\,\ref{tab:filament:calculated}) relates to $M_{\rm F}/(\pi r_{\rm F}^2 L)$, where $r_{\rm F}$ represents the radius of the filament ($r_{\rm F} = \frac{1}{2} W_{\rm F}$).

Within the footprint of the IRS\,17 filament, ten 3\,mm dense cores are identified (see Fig.\,\ref{fig:band3:con}). These cores were identified by \cite{2021MNRAS.505.2801L}, who utilized both the {\it astrodendro} algorithm and the {\it CASA imfit} function together for the identification. These cores have a mass range of 0.3-2.5\,\msun\, (see Table\,4 of \citealt{2021MNRAS.505.2801L}) with sizes of the order of 0.01\,pc. However, one of the cores (i.e., \#2, see Fig.\,\ref{fig:band3:con}), with a mass of 2.0\,\msun, is associated with the ionized gas excited by the intermediate-mass IRS\,17 protocluster. Therefore, its mass could
be overestimated by assuming that 3-mm continuum emission arises
primarily from dust thermal emission. As such, the mass of this core should be interpreted with caution. It is worth noting that the same issue could influence the mass calculation of the natal filament, but not significantly. This is because the absolute value of the mass of the related core, which encompasses most of the ionized gas, is much lower than the filament mass. 
Furthermore, these dense cores within the filament do not exhibit a periodic chain-like distribution as seen in Fig.\,\ref{fig:band3:con}, which will be discussed in Sect.\,\ref{sub:fiber:star:formation}.

\subsection{Fibers and condensations in QUARKS 1.3-mm dust continuum }
\label{subsubsec:band6:cont}

Figure\,\ref{fig:bad6} displays the QUARKS 1.3-mm dust continuum map of the IRS\,17 protocluster region (also refer to Fig.\,\ref{fig:band3:band6}b). 
The IRS\,17 filament can still be observed in 1.3-mm emission, confirming the filamentary nature seen in 3-mm emission. This filament presents several substructures in 1.3-mm observations, including small-scale fibers and dense condensations. To identify them, the {\it getsf} algorithm was applied to the 1.3-mm dust continuum map. 
The bona fide fibers were identified based on the following three criteria: i) the peak flux of a fiber is detected at least at 3\,$rms$ significance level, ii) the length of a fiber is at least 3 times the synthesized beam size at 1.3-mm, and iii) the aspect ratio is at least 3.

As a result, a total of 7 fibers and 24 condensations were obtained (Fig.\,\ref{fig:bad6}). However, upon  a careful visual inspection,
several dense fibers and condensations appear not be recognized by the {\it getsf} algorithm. Specifically, a network of fibers between condensations 25, 28 and 29 is  seen in the figure but not detected by {\it getsf} (see Fig.\,\ref{fig:band6:zoomin}). This is because they are disjoint due to low peak emission of $<3$\,$rms$ along each individual fiber in the network. In addition, this network of fibers does not appear as velocity coherent structures, which can be found in the \htcop~(1--0) velocity field map (see Fig.\,\ref{fig:h13co:moment}b). To recover the missing condensations, another source extraction method, SExtractor \citep{1996A&AS..117..393B,2016zndo....804967B} was utilized, as adopted in other studies (e.g.; \citealt{2013A&A...549A..45C,2014A&A...568A..41U,2024arXiv240402275X}). This approach identified 6 additional condensations (i.e., C\,1, C\,18, C\,19, C\,22, C\,23, and C\,24), leading to a total of 30 condensations (refer to Table\,\ref{tab:core}). Except for C\,30, all condensations are located in the IRS\,17 filament (see Fig.\,3). 
The parameters measured by the {\it getsf} for fibers are listed in Table\,\ref{tab:filament}, including the center positions, length, width, and integrated flux. The parameters by the {\it getsf}  or the SExtractor for condensations are given in Table\,\ref{tab:core}, including the center positions, size, integrated flux and peak flux.

The gas kinematics of the IRS\,17 filament and the identified fibers was inferred using the \htcop~(1--0) molecular line transition from the ATOMS survey.
The velocity centroid and the velocity dispersion of both filament and the fibers are tabulated in Table\,\ref{tab:filament:calculated}.

\subsubsection{Derived parameters of fibers and condensations}\label{subsub:sec:fibers and condensation}
Following Eq.\,\ref{equ:mass:cal}, we quantified the mass of fibers ($M_{\rm f}$) and condensations ($M_{\rm c}$) from the integrated flux at 1.3-mm, given the corresponding dust opacity $\kappa_{\rm 1.3mm} = 0.899\,\mathrm{cm^2\,g^{-1}}$ \citep{1994A&A...291..943O} and the temperature of 25\,K. Using the same approach as for the IRS\,17 filament, we can estimate
for fibers the line mass ($m_{\rm f}$), and the average number density ($\langle n_{\rm f} \rangle$).
For condensations, assuming a uniform spherical density distribution, their number density can be inferred as $n_{\rm c} = \frac{3M_{\rm c}}{4\pi R_{\rm c}^3 \mu m_H}$, the mean molecular weight $\mu=2.8$, and $m_H$ is the mass of the hydrogen atom. $R_{\rm c}$ is the radius of each condensation, given by $R_{\rm c} = \sqrt{R_{eff}^2-(1/4\,\ln{2}){\rm bmaj}\times\,{\rm bmin}}$ \citep{2014ApJ...790...84L}, where the effective radius, $R_{eff}= \sqrt{FWHM_{\rm maj} \times FWHM_{\rm min}}/2$).

Table\,\ref{tab:filament:calculated} and Table\,\ref{tab:core} outline the parameters derived above for the fibers and condensations, respectively. The fibers exhibit a length range of  $\sim3000$ to $9000$\,au, with a median value of $\sim4500$\,au. Their width varies from about $\sim1000$ to $2000$\,au, with a median of roughly 1400\,au. The mass of these fibers spans from 0.66 to 2.72\,\msun, with a median of $\sim1.2$\,\msun. The line mass ($m_{\rm f}$) extends from 31.4 to 82.0\,\msun\,${\rm pc^{-1}}$, with a median of 51.9\,\msun\,${\rm pc^{-1}}$. The average number density of fibers varies within a range of $\rm 1.3\times10^7-6.2\times10^7\,cm^{-3}$  with a median of ${\rm 1.8\times10^7\,cm^{-3}}$.
As for the condensations, their effective radius and number density have a range of [180, 410]\,au, and [$7.3\times10^7$, $2.7\times10^9$]\, ${\rm cm^{-3}}$, respectively. The mass of these condensations spans from 0.02 to 1.31\,\msun, with a median of 0.11\,\msun. These results for condensations are comparable to values derived in other intermediate-mass star-forming regions, such as the condensation mass ranges from 0.1 to 0.4\,\msun\, in the TUKH122 prestellar core in the Orion A cloud \citep{2018ApJ...856..147O}, and the mass ranges from 0.01 to 0.2\,\msun\, in the NGC 2071 IR \citep{2022ApJ...933..178C}.
It is worth noting here that the derived masses could be underestimated for two likely reasons (1) the assumed average clump temperature of 25~K could be on the higher side for some of the non-YSO bearing condensations and (2) the high-resolution interferometric observations could have filtered out diffuse, low-density emission. Additionally, we cannot rule out over estimation of mass for condensations harboring protostars where temperatures could be as high as 50\,K.

\subsubsection{Classification of fertile fibers}

As depicted in Fig.\,\ref{fig:bad6}, some fibers accommodate dense condensations, which are likely precursors to star formation. Given their spatial association with condensations, fibers can be classified as ``fertile" or ``sterile" \citep{2013A&A...554A..55H,2015A&A...574A.104T}. 
According to these authors,fertile fibers generally have a higher line mass and contain at least three dense condensations, unlike their sterile counterparts. In the IRS\,17 filament, four fibers (namely, f\,1 and f\,5--7) 
are classified as fertile, while the rest are considered sterile.

We applied the minimum-spanning tree (MST) method \citep{1985MNRAS.216...17B,2016ApJS..226....9W,2019ApJ...886..102S,2022MNRAS.510.5009L,2024ApJS..270....9X} to estimate the separation between condensations within the fertile fibers.  
The MST approach (e.g. \citealt{2019A&A...629A.135D}) determines the shortest distances that can possibly connect each of the objects in a given field. 
The estimated separations of condensations in the four fertile fibers range from approximately 850 to 3450\,au, with a median of around ${\rm 1600\pm600\,au}$, where the error is the standard deviation of their separations.

\subsection{Filament and fibers seen in molecular gas emission}\label{sub:fil:fib:gas}
\begin{figure}[ht!]
    \centering
    \includegraphics[angle=0, width=0.47\textwidth]{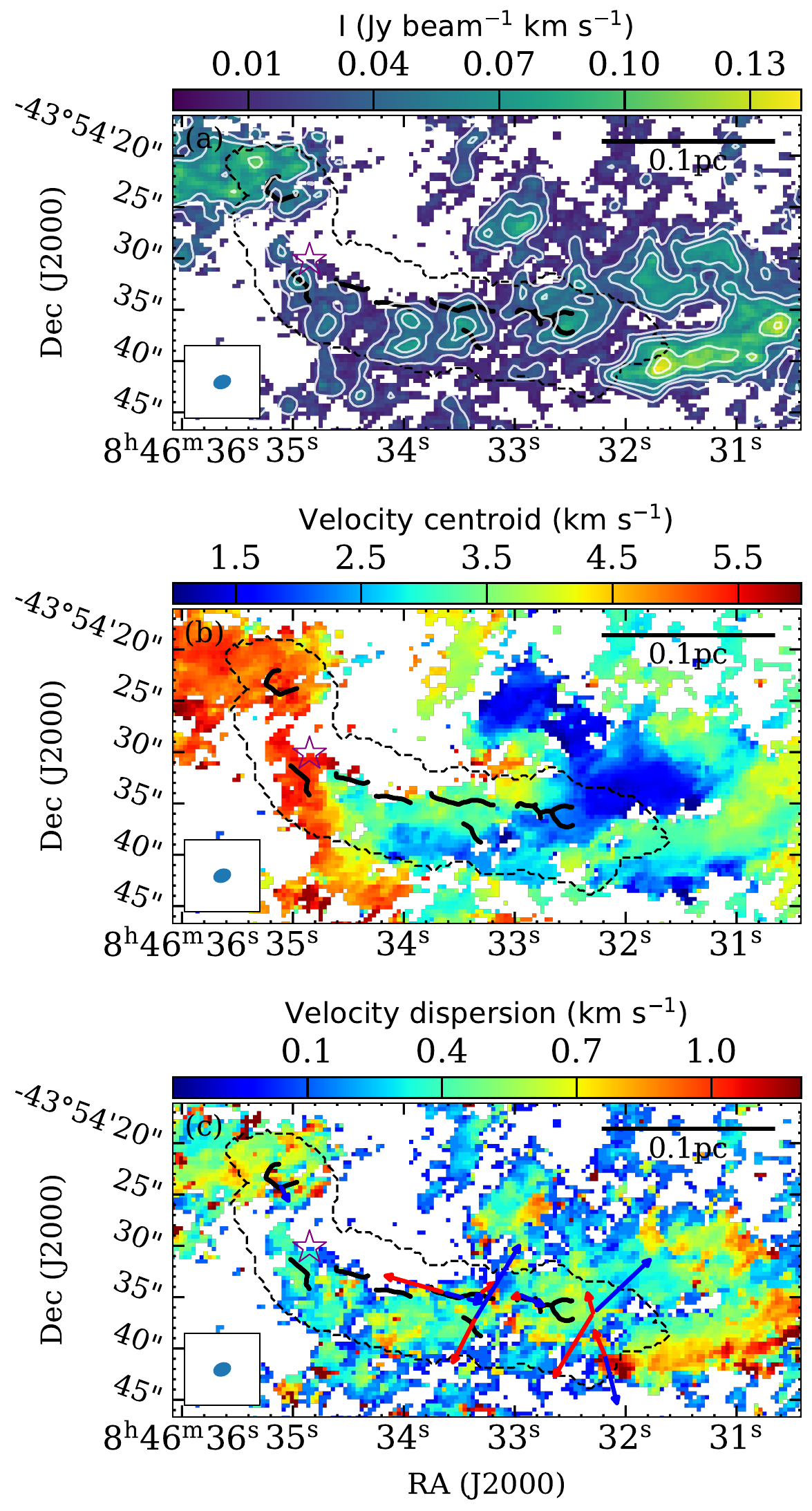} 
    \caption{Moment maps of \htcop\, from the ATOMS survey. The velocity range of \htcop\, emission is set to [1, 6]\,$\rm km\,s^{-1}$, which correspond to the velocity channels with a signal-to-noise ratio above 3 $rms$ in the average spectrum over the entire region. The $rms$ defined as the standard deviation within 50 channels free of  line emission ($\sim0.01\,\rm Jy\,beam^{-1}$). The black dashed and solid lines represent the IRS\,17 filament footprint and fibers, respectively. The ATOMS beam
    and 0.1\,pc scale-bar are shown in left bottom and top right, respectively. (a) Integrated intensity (Moment\,0) map. The white contour levels start at 3 $rms$ ($\sim\rm 0.01\,Jy\,beam\,km\,s^{-1}$), and following as [6, 9, 12, 15]\,$rms$. (b) Velocity centroid (Moment\,1) map. (c) Velocity dispersion (Moment\,2) map. The solid and dashed arrows represent the outflows identified either from CO (2--1) or SiO (5--4), respectively (see Fig.\,\ref{fig:outflow})} 
    \label{fig:h13co:moment}    
\end{figure}

Figure \ref{fig:h13co:moment} presents the moment maps of \htcop\,(1--0) toward the IRS\,17 filament observed in 3-mm dust continuum. The \htcop\,(1--0) line, with a critical density of $\sim\rm 10^{5}\,cm^{-3}$ \citep{2012ApJ...756...60S}, allows us to trace relatively dense gas compared to CO and its isotopologue lines \citep{2023A&A...672A.133S}. The Moment\,0 map of \htcop\,(1--0) reveals that relatively dense gas emission spans the entire filament, except for the region around the intermediate-mass protocluster (see the purple star symbol in Fig.\,\ref{fig:h13co:moment}) where protostellar feedback is supposed to be strong. In addition, near the protocluster, 
\htcop\,(1--0) emission does not cover the full filament footprint. 

This could be attributed to the missing flux in the interferometric data or to the clean-up by the outflows with a wide opening angle from the IRS\,17 protocluster (see the red-shifted high-velocity CO gas component related to the protocluster in Fig.\,\ref{fig:outflow}a). 
As a consequence, though the \htcop\,emission appears to be primarily distributed south of the crest of the IRS\,17 filament, particularly in the southwest region, the morphology and the location of identified fibers indicate that it traces the 3-mm filament.

In the velocity field (Moment\,1) map of \htcop, the molecular gas shows a coherent velocity distribution along the dust emission filament.  Similar to the velocity structure of the filamentary cloud in Taurus, presented in \citet{2013A&A...554A..55H}, multiple velocity components are seen in the IRS\,17 filament, ranging from 1.5 to 5.5$\rm \,km\,s^{-1}$ (see \htcop\ velocity field in Fig\,\ref{fig:h13co:moment}b), suggesting complex gas structures. Here, these velocity components can also be observed in the velocity map of the DCN line (see Fig.\,\ref{fig:dcn:moment1}) from the QUARKS survey at higher angular resolution but lower detection rate than the same map of the \htcop\ line from the ATOMS survey. 
Additionally, a distinct velocity gradient is also observed along the filament, with increasing velocity centroid from west to east. Despite this complex gas kinematics observed in the filament, the individual identified fibers are seen to be velocity coherent sub-structures (the velocity centroid ranges from 2.3 to 5.3\,$\rm\,km\,s^{-1}$, see Col.\,5 in Table\,\ref{tab:filament:calculated}). Furthermore, the detected fibers are seen to be in three bundles with similar kinematics, (1) f\,1--2 corresponding to a velocity of $\sim$5.0\,$\rm km\,s^{-1}$, (2) f\,3--6 with velocity $\sim$3.1\,$\rm km\,s^{-1}$, and (3) f\,7 with velocity $\sim$2.3\,$\rm km\,s^{-1}$.

In Fig.\,\ref{fig:h13co:moment}c, we present the velocity dispersion (Moment\,2) map of \htcop~(1--0). Within the IRS\,17 dust filament, molecular gas has a median velocity dispersion of $0.39\pm0.25\,\rm km\,s^{-1}$, indicating transonic gas motions compared to the typical thermal sound speed of 0.3\,$\rm km\,s^{-1}$ at a temperature of 25\,K. The locally enhanced gas motions likely result from local active star formation feedback, including strong ionized gas from the intermediate-mass IRS\,17 protocluster (see the purple star symbol in Fig.\,\ref{fig:h13co:moment}c) and outflows (see the arrows in Fig.\,\ref{fig:h13co:moment}c) distributed along the filament. Specifically, the northeastern region near the protocluster bears more pronounced supersonic motions due to ionized gas feedback, while several outflows scattered along the filament make local gas motions enhanced. Additionally, several regions outside the filament appear even higher supersonic velocities (i.e., $\rm \sim 1\,km\,s^{-1}$), which deserves further investigation in the future.
\cite{2018A&A...610A..77H} investigated the massive Integral Shape Filament (ISF) in Orion, which experiences more intense star formation feedback. Their observations of the ISF fibers using the $\rm N_2H^+$~(1-0) transition revealed transonic non-thermal velocity dispersions (0.24\,\vel\ on average) relative to the local temperature. Given this, the velocity dispersion estimated from \htcop~(1–0) appears reasonable for the IRS\,17 region.

\subsection{Stability of the IRS\,17 filament and fibers}
\label{subsub:instability}

The gravitational stability of filaments and 
fibers can be quantitatively evaluated under an ideal model of isothermal and infinite cylinders using the parameter of the critical line mass ($m_{\rm crit}$, \citealt{1963AcA....13...30S,1964ApJ...140.1056O}). 
Filaments and fibers with an observed line mass greater than $m_{\rm crit}$ (supercritical) are expected to collapse under their own gravity, while those with an observed line mass less than $m_{\rm crit}$ (subcritical) can maintain a hydrostatic equilibrium or become dispersed.

The $m_{\rm crit}$ parameter can be expressed as follows \citep{1963AcA....13...30S,1964ApJ...140.1056O}:
\begin{equation}\label{eq:cseff:cri}
    m_{\rm crit}=\frac{2\sigma^2_{\rm eff}}{G},
\end{equation}
where $G$ represents the gravitational constant and $\sigma_{\rm eff}$ is the effective speed of sound to be considered in varied situations.
If we consider only the contribution from thermal motions, the term $\sigma_{\rm eff}$ can be replaced with the thermal sound speed of gas, $c_{\rm s}$, resulting in the thermal critical line mass:
\begin{equation}\label{eq:th:cri}
    m_{\rm crit}^{\rm th}=\frac{2c^2_{\rm s}}{G} \sim 16.6(\frac{T}{\rm 10\,K})M_{\odot}\,\rm pc^{-1}\,\rm ,
\end{equation}
where $T$ is the kinetic temperature of the gas. Here, we assume it to be the same as the average dust temperature of $T=$ 25\,K. This yields $m_{\rm crit}\sim$ 41.5\,\msun\,$\rm pc^{-1}$. Compared with the observed line mass, the IRS\,17 filament and all the fibers except f\,2 and f\,4 are thermally supercritical (see Table.\,\ref{tab:filament:calculated}).  It is important to note that the assumed average dust temperature of 25\,K is relatively high compared to the estimated temperatures of other intermediate-mass protoclusters, such as NGC 1333 ($\sim$10\,K; see Table 2 of \citealt{2018A&A...610A..77H}). However, adopting a temperature of 10\,K would render fibers f\,2 and f\,4 supercritical, which contradicts their observed nature since no associated condensations are present (see Fig.\,\ref{fig:band3:con}). Therefore, accurate temperature measurements will be essential in future studies to resolve this discrepancy.

If we account for the contribution from non-thermal motions to the stability of the IRS\,17 fibers, the term $\sigma_{\rm eff}$ can be substituted with the total velocity dispersion, $\sigma_{\rm tot}$, and thus the turbulent critical line mass becomes
\begin{equation}\label{eq:tur:cri}
    m_{\rm crit}^{\rm tur}=\frac{2\sigma_{\rm tot}^2}{G} \sim 465(\frac{\sigma_{\rm tot}}{\rm 1\,km\,s^{-1}})^{2}M_{\odot}\,\rm pc^{-1}.
\end{equation}
From the \htcop~(1--0) line data, we obtained an median velocity dispersion of $\sim 0.39\pm 0.25\,{\rm km\,s^{-1}}$ within the IRS\,17 filament (see Table\,\ref{tab:filament:calculated}). 
Using this, we estimate a turbulent $m_{\rm crit}^{\rm tur}$ of $70.7\pm 29.1$\,$M_{\odot}\,\rm pc^{-1}$. 
Given the complex velocity field and within the quoted uncertainties, the IRS\,17 filament could be supercritical, consistent with the fragementation to fibers hypothesis discussed in Section \ref{sub:fiber:star:formation}. However, the estimated turbulent $m_{\rm crit}^{\rm tur}$ is significantly greater than the observed line mass of fibers within IRS\,17 region (except for f\,3, see Col.3 in Table.\,\ref{tab:filament:calculated}), with a factor of 1.1 to 2.3. 
This contradicts the observed scenario of gravitational collapse of the fibers and the formation of the detected condensations within them. The discrepancy is likely produced by the ongoing active star formation within IRS\,17 filament, where outflows and feedback from ionized gas results in large velocity dispersion. Such high turbulence may not be prevalent during the initial stages of collapse of the small-scale fibers, but is typically close to the sound speed \citep{2013A&A...554A..55H,2018A&A...610A..77H}.
The above discussion finds support in the observed 
cascade of turbulence with the decreasing scale of the density structures \citep{1981MNRAS.194..809L,2023ASPC..534..153H,2024AJ....167..228L}. The $\sigma_{\rm tot}$ follows $\sigma-L$ ($\sigma_{\rm tot}=c_{\rm s}(1+\frac{L}{\rm 0.5\,pc})^{0.5})$, an empirical relation observed by \cite{2023ASPC..534..153H} for filaments and fibers, which is akin to the first of Larson's relation for general clouds \citep{1981MNRAS.194..809L}. 
For the  estimated length scale of the IRS\,17 fibers, which are significantly less than 0.05\,pc ($\sim$ 10000\,au), $\sigma_{\rm tot} \sim c_{\rm s}$ making $m_{\rm crit}^{\rm tur} \sim m_{\rm crit}^{\rm th}$. Thus, for the length scale of the IRS\,17 fibers, the role of turbulence would be negligible. Summarizing the stability analysis, except f\,2 and f\,4, the other IRS\,17 fibers are observed to be supercritical consistent with the presence of dense condensations within them.
However, we should note that the $\sigma-L$ has uncertainties of $\pm0.2\,\rm pc$ and $\pm0.2$ for the length ($L$) and exponent, respectively \citep{2023ASPC..534..153H}. 

Note that magnetic fields could potentially have a considerable impact on the stability of the fibers. However, our analysis above ignores the magnetic fields, as such information is not available.

\begin{figure}[ht!]
    \centering
    \includegraphics[angle=0, width=0.45\textwidth]{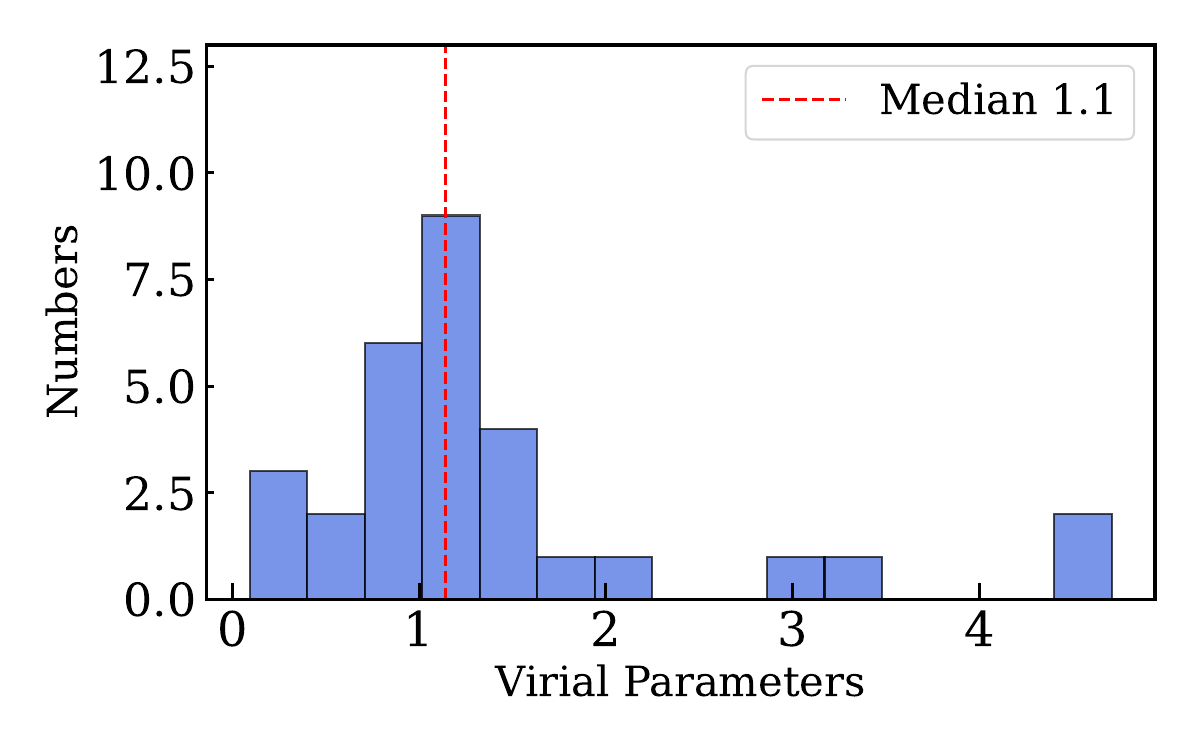} 
    \caption{Virial parameter distribution of IRS\,17 condensations. Only the thermal motion is considered for virial analysis. The red dashed lines represent the median value of virial parameter.} 
    \label{fig:vir}    
\end{figure}
\subsection{Stability of condensations}
The stability of condensations can be evaluated by the virial parameter ($\alpha_{\rm vir
}$) as below \citep{1992ApJ...395..140B}:
\begin{equation}
    \alpha_{\rm vir}=\frac{5\sigma_{\rm eff}^2R_{c}}{\rm G{\it M_{\rm c}}}\,,
\end{equation}
where $R_{\rm c}$ is the condensation radius. The subvirial cores with $\alpha_{\rm vir}\leq2$ will collapse to form stars \citep{2022MNRAS.510.5009L}. 
For the stability analysis of the condensations, one needs to include the contribution from turbulence in the $\sigma_{\rm eff}$ term. However, we lack molecular line data (e.g., $\rm N_2H^+$~(1--0)) at matching angular resolution. 
Moreover, if IRS\,17 fibers exhibit transonic turbulent motions similar to the OrionA ISF fibers as reported by \cite{2018A&A...610A..77H} using $\rm N_2H^+$ observations, the impact of turbulence on the stability of smaller-scale condensations will become negligible. This is consistent with $\sigma - L$ scaling (see Section \ref{subsub:instability}), where the effect of turbulence would be negligible for the length scales of these sub-structures. Hence, we consider only the thermal motion and replace
$\sigma_{\rm eff}$ by the sound speed of gas, $c_{\rm s}$. Assuming the same $c_{\rm s}$ value (i.e., $0.3\,\rm km\,s^{-1}$) for all condensations in the IRS\,17 filament at a temperature of 25\,K, we find 
most (25 of 30, $83\,\%$) of them with $\alpha_{\rm vir}\leq 2$, with a median value of 1.1 (see Fig.\,\ref{fig:vir}). 
This implies that most of identified condensations within the IRS\,17 region are gravitationally bound and will undergo collapse unless supported by magnetic field. 
Moreover, two of the condensations (C5 and C21) are associated with the Class\,I/0 YSOs (see Fig.\,\ref{fig:bad6}), and additional seven (i.e., C2, C10, C13, C16, C17, C25 and C26) are associated with outflows identified from the SiO and CO line transitions (see Fig.\,\ref{fig:outflow}). The association of these YSOs and outflows lends strong support to our understanding that these condensations are potential seeds of star formation.
It is worth noting that, as mentioned in Section \ref{subsub:sec:fibers and condensation}, the condensations without YSO associated could have lower temperatures possibly down to 10K, while those with YSO associated could have higher temperature possible up to 50 K. The former case will make the virial parameters about 6 times lower than the current values, and thus does not change the subvirial state of those condensations without YSOs; the latter case will make the virial parameters about 4 times higher, and accordingly make the dynamical state of those condensations with YSOs change to be supervirial.

A point that solicits further discussion is the low masses of the identified condensations (see Table\,\ref{tab:core}). Notwithstanding the issue of assuming a higher dust temperature  or factoring in the missing flux of the interferometric observations, for the non-YSO bearing and non-outflow hosting condensations, the estimated masses are similar to those derived for the protostars in NGC 2071 IR by \citet{2022ApJ...933..178C} using 1.3~mm ALMA observations. As pointed out by these authors, these could be treated as lower limits, since the emission from the protostars could be partially optically thick at 1.3~mm. Furthermore, with the sensitivity and the resolution of the observed data, these detected supercritical ($\alpha_{\rm vir}\leq 2$) condensations could be just the central cores of Bonnor–Ebert spheres that have steep density profiles \citep{2018ApJ...856..147O}. However, one cannot rule out the formation of very low-mass objects from these condensations. The condensations associated with YSOs and outflows, on the other hand, might be undergoing a dynamical process of star formation, as described in various dynamical models of star formation (e.g., \citealt{2019MNRAS.490.3061V,2020ApJ...900...82P}). For instance, the inertial inflow model proposed by \cite{2020ApJ...900...82P} outlines star formation as a three-step dynamical process: (1) the formation of a gravitationally unstable core, (2) the collapse of this core into a low to intermediate-mass star, and (3) the subsequent accretion of the remaining mass from the natal core onto the star. An exception is the condensation C21, where the estimated mass is too low ($0.02\pm0.01$\,\msun\,) to explain the signature of star formation therein. Given the weak 1.3~mm emission observed, it is highly likely that we are detecting only the peak emission from the central region.

\begin{deluxetable*}{lcccccc}
\tabletypesize{\footnotesize}
\tablecaption{Derived parameters of fibers \label{tab:fibers}}
\tablewidth{0pt}
\tablehead{
    Fiber      &  \multicolumn{2}{c}{\underline {~~~~~~~~~~Position~~~~~~~~~~}} &$\langle n_0\rangle^a$ &  $\rm <\lambda_{\rm pro}>^b$ & H &  $ \lambda_{\rm max}$  \\
              & \colhead{$\alpha$(J2000)} & \colhead{$\delta$(J2000)}& $(\rm \times10^8\, cm^{-3})$&  $(\rm \times 10^3\,au)$ & $(\rm \times 10^2\,au)$ & $(\rm \times 10^3\,au)$  }
\startdata
    f1     &08:46:35.12  & -43:54:23.44&  $2.18\pm0.34$  &  $2.3\pm0.4$ & $0.7\pm0.1$ & $1.8\pm0.1$  \\
    f5     & 08:46:33.49  & -43:54:34.66&   $3.18\pm0.63$  &  $1.9\pm1.0$  &  $0.6\pm0.1$ & $1.2\pm0.2$ \\
    f6     & 08:46:33.40  & -43:54:38.20& $8.97\pm1.42$  &  $0.9\pm0.2$ &$0.3\pm0.1$ &   $0.7\pm0.1$  \\
    f7     &08:46:32.69  & -43:54:36.17&  $2.21\pm0.34$  &  $1.7\pm0.2$  & $0.7\pm0.1$   & $1.5\pm0.3$  \\
\enddata
\begin{flushleft}
$^a$ The central density is estimated from the average volume density of condensations within each fiber.\\
$^b$ The mean separation between condensations of each fiber, and the errors from the standard deviation of the separations.. 
\end{flushleft}
\end{deluxetable*}
\subsection{Fragmentation of fibers}\label{sub:frag}

Theoretically, semi-analytic linear cylinder fragmentation models, such as the “sausage” instability model \citep{1987PThPh..77..635N,2010ApJ...719L.185J}, are commonly used to investigate the fragmentation of filamentary structures. The premise of these theoretical models is that the fragmentation of a self-gravitating fluid cylinder can lead to chains of dense cores with nearly periodic separations \egcite{1953ApJ...118..116C,1987PThPh..77..635N,2010ApJ...719L.185J}. This separation corresponds to the characteristic wavelength ($\lambda_{\rm crit}$) where the instability grows most rapidly.  For an incompressible fluid, perturbation analysis indicates that the critical wavelength emerges at $\lambda_{\rm crit} =  11 R_{\rm cyl}$, where $R_{\rm cyl}$
 is the cylinder's radius \egcite{1953ApJ...118..116C}. 
In the context of an infinite isothermal gas cylinder, the characteristic wavelength becomes $\lambda_{\rm crit} = 22 H$, where $H = c_{\rm s} (4\pi G \rho_{\rm c})^{-1/2}$ represents the isothermal scale height and $\rho_{\rm c}$ is the central mass density along the cylinder's axis \egcite{1987PThPh..77..635N,1992ApJ...388..392I,2010ApJ...719L.185J}.
For a finite isothermal gas cylinder surrounded by a uniform medium, the value of $\lambda_{\rm crit}$ is influenced by the $R_{\rm cyl}/H$ ratio. If $R_{\rm cyl}/H \gg 1$, the characteristic separation approximates $\lambda_{\rm crit} = 22 H$, aligning with the infinite isothermal gas cylinder model. However, if $R_{\rm cyl}/H \ll 1$, it shifts to $\lambda_{\rm crit} = 11 R_{\rm cyl}$, which corresponds to the incompressible case \egcite{2010ApJ...719L.185J}.

For the IRS\,17 filament, we observe no chain of dense cores with a characteristic periodic separation. This suggests that the IRS\,17 filament fragmentation into dense cores does not conform to predictions of a simple linear cylinder model. In comparison, predictions of the model as
patterns of periodically-spaced dense condensations are evident in the fertile fibers (i.e., f\,1, and f\,5--7) within the filament. Hence, assuming the fibers as isothermal and hydrostatic gas cylinders, we take the average volume density of condensations (refer to Col.4 in Table.\,\ref{tab:fibers}) within each fiber as its central density ($\rho_{\rm c}$). This allows us to estimate the isothermal scale height for the fibers, yielding $H\sim60\,\rm au$ at a temperature of 25\,K. The median radius of the fertile fibers is $\sim$700\,au ($1/2\,\times W_{\rm f}$), which corresponds to the case of $R\gg H$. Hence, the separation of dense condensations within fibers can be predicted from $\lambda_{\rm max}=22H$. Consequently, the predicted $\lambda_{\rm max}$ for the four fertile fibers  ranges from $700-1800\,\rm au$ (see Col.\,7 in table\,\ref{tab:fibers}). This agrees with the observed typical separation between dense condensations within fertile fibers of $\sim 900-2300\rm\,au$ (see Col.5\, in table\,\ref{tab:fibers}).

\section{Discussion}\label{sec:dsicussion}
\subsection{Role of fibers in star formation}\label{sub:fiber:star:formation}
\begin{figure}[ht!]
    \centering
    \includegraphics[angle=0, width=0.45\textwidth]{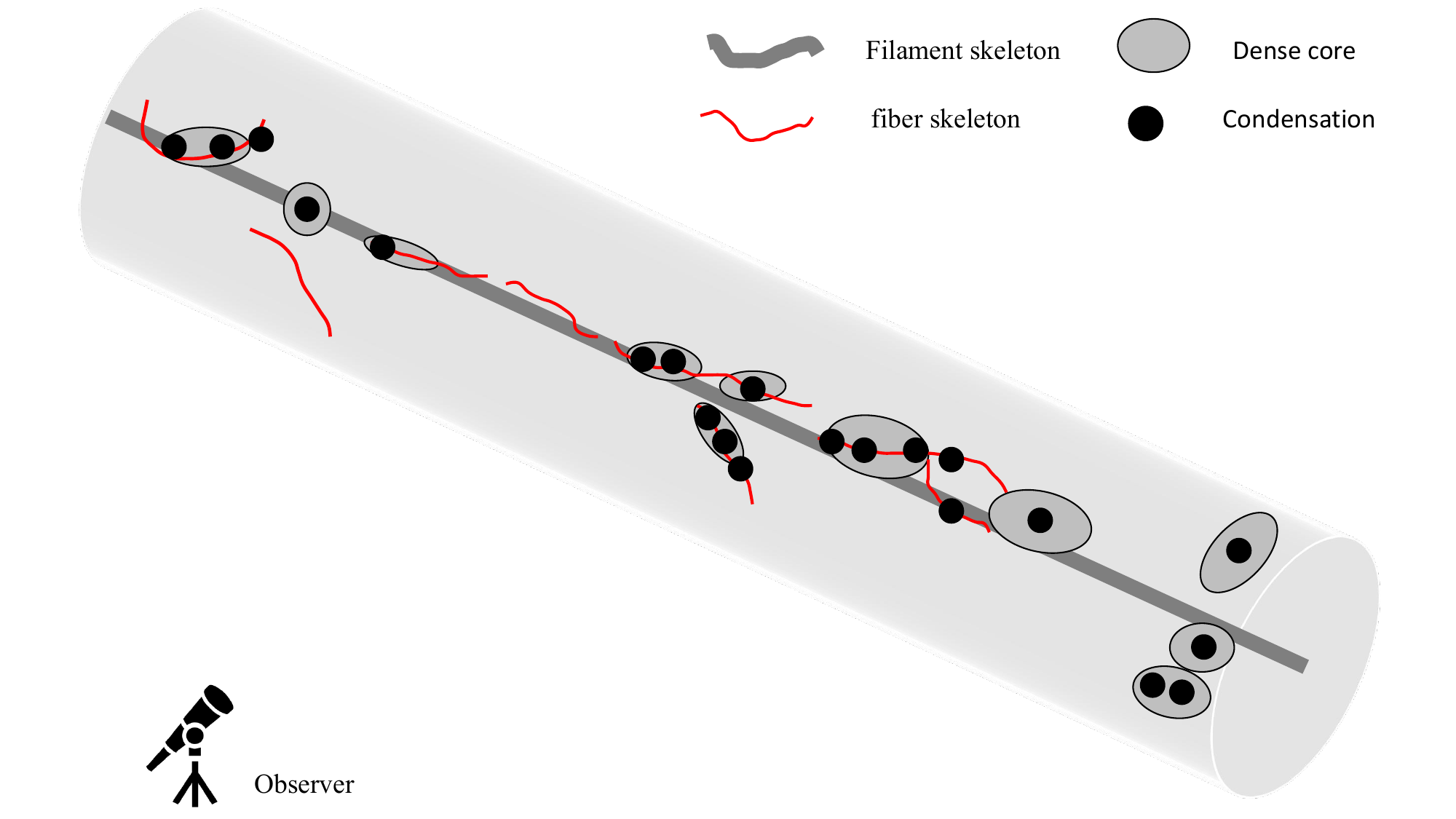} 
    \caption{Cartoon of a cylinder filament model.}
    \label{fig:sheet:cylinder:model}    
\end{figure}
Since their discovery \citep{2013A&A...554A..55H},
subparsec-scale fibers have been reported across 
varied star-forming regions, including low-mass clouds \citep{2013A&A...553A.119A,2016A&A...590A..75F}, intermediate-mass clusters \citep{2014ApJ...790L..19F,2017A&A...606A.123H}, and high-mass star-forming regions \citep{2018A&A...610A..77H,2019A&A...629A..81T,2022ApJ...926..165L}. In good agreement with observations, hydrodynamical simulations of turbulent clouds reveal the natural occurrence of network of fibers within filaments \citep{2014MNRAS.445.2900S,2016MNRAS.455.3640S}. 
Furthermore, in most regions, the population of observed cores are seen to be predominantly harboured in these fibers, thus highlighting the key role played by these dominant sub-structures in star formation across the mass spectrum. Such a scenario is discerned in the IRS\,17 filament, with the observed association between fibers and dense condensations (potential seeds of star formation).
A systematic analysis of the internal fibers of the massive Orion-ISF filament and lower-mass objects (e.g., Musca, Taurus) led \cite{2018A&A...610A..77H} to propose
a unified star formation paradigm. In this framework,
the observed differences between low- and high-mass clouds naturally emerge from the initial concentration of fibers. 

Fibers are proposed to play a critical role in star formation. \citet{2013A&A...554A..55H} and \citet{2015A&A...574A.104T} based on their observations of L1495/B213 filamentary complex in Taurus, infer a hierarchical fragmentation process termed `fray and fragment', where the filamentary clouds first fragment to fibers (sub-filaments) which further fragment to dense cores. The above scenario of the impact of fibers on the fragmentation of filaments finds strong support in the numerical study discussed in \citet{2020MNRAS.497.4390C}. 
 
As shown by these authors, in contrast to the classical filament fragmentation models, the presence of fibers as the intermediate fragmentation stage results in the absence of a strong imprint of quasi-periodic cores in filaments. 

The above picture is aptly captured in the cartoon presented in  Fig.\,\ref{fig:sheet:cylinder:model}, which explains the observed features in the IRS\,17 filament. As discussed in Section \ref{sub:fil:fib:gas}, the IRS\,17 filament presents 
multiple velocity components, consistent with the three bundles of velocity-coherent fibers identified. The detected cores are randomly spaced in the large-scale filament. This is in good agreement with recent dynamic models using numerical simulations that consider accretion, turbulence, and magnetic fields to interpret the observed randomly-distributed core spacings \citep{2015MNRAS.452.2410S,2016MNRAS.458..319C,2017MNRAS.468.2489C,2020MNRAS.497.4390C,2017ApJ...834..202G}. However, the signature of periodic core-spacing is observed in a few of the detected velocity-coherent, fertile fibers in IRS\,17 filament. Not consistent with the above picture of hierarchical fragmentation, are the hydrodynamical simulations by \citet{2014MNRAS.445.2900S,2016MNRAS.455.3640S}. These authors propose a `bottom-up' model, where fibers pre-exist in clouds, formed as a natural outcome of a turbulent cascade. Subsequently, these fibers are swept together to form the large-filaments as a consequence of the large-scale gravity-driven collapse of the cloud. The differing views advocate for further such studies of filamentary clouds at higher resolution and sensitivity for a better insight into the formation and role of fibers in star formation.

\section{Summary}\label{sec:summary}
We have investigated the internal structure of the IRS\,17 filament in the intermediate mass protocluster, IRAS 08448-4343, using both 3\,mm and 1.3\,mm ALMA data from the ATOMS and QUARKS survey, respectively. Our study has illustrated the role of small-scale fibers in star formation.  Our major results are as follows:

A sub-parsec filament ($\sim$0.26\,pc), referred to as the IRS\,17 filament, is seen in 3\,mm continuum emission  within the IRS\,17 protocluster region. The gas kinematics of this filament presents a complex and multi-component field.

Within the apparent IRS\,17 filament, three bundles of seven  velocity-coherent fibers and twenty-nine dense condensations were revealed from the QUARKS 1.3\,mm continuum map. The fibers have a median length of $\sim 4500\,\rm au$ and a median width of $\sim 1400\,\rm au$ with a line mass of 51.9\,\msun $\rm ~pc^{-1}$. We identified four fertile fibers, each hosting three or more condensations, based on the association between condensations and fibers. Some condensations are linked to ongoing star formation signatures, such as YSOs and outflows, while most are subvirial. These findings suggest that the majority of condensations could potentially seed star formation.

Furthermore, periodic spacings of condensations are observed in the fertile fibers, which agrees with linear isothermal cylinder fragmentation models. However, this periodic pattern is not observed for the detected 3\,mm cores in the larger-scale IRS\,17 filament.
This reflects the crucial role of fibers on the fragmentation of the filament into cores. In other words, there may be an intermediate fragmentation step, involving the formation of fibers, before core formation out of the IRS\,17 filament. This step leaves no strong imprint of quasi-periodic fragmentation of filaments into cores.

\section*{Acknowledgements}
We thank the referee for the valuable comments and suggestions that greatly improved the quality of this paper. This work has been supported by the National Key R\&D Program of China (No.\,2022YFA1603101).
H.-L. Liu is supported by National Natural Science Foundation of China (NSFC) through the grant No.12103045, and by Yunnan Fundamental Research Project (grant No.\,202301AT070118, 202401AS070121).
T. Liu acknowledges the supports by NSFC through grants No.12073061 and No.12122307; and X. Liu has also been supported by CPSF No.\,2022M723278; S.-L. Qin is supported by NSFC under No.12033005.
G.G. gratefully acknowledges support by the ANID BASAL project FB210003.
PS was partially supported by a Grant-in-Aid for Scientific Research (KAKENHI Number JP22H01271 and JP23H01221) of JSPS, and by Yoshinori Ohsumi Fund (Yoshinori Ohsumi Award for Fundamental Research).
This work was performed in part at the Jet Propulsion Laboratory, California Institute of Technology, under contract with the National Aeronautics and Space Administration (80NM0018D0004).
SRD acknowledges support from the Fondecyt Postdoctoral fellowship (project code 3220162) and ANID BASAL project FB210003. This work was partly supported by Yunnan Xingdian Talent Support Plan–Youth Project. LB gratefully acknowledges support by the ANID BASAL project FB210003.
This paper makes use of the following ALMA data: ADS/JAO.ALMA\#2019.1.00685.S, 2021.1.00095.S and 2023.1.00425. 



\appendix
\section{Complementary figures and tables}

\begin{figure*}[ht!]
    \centering
    \includegraphics[angle=0, width=0.8\textwidth]{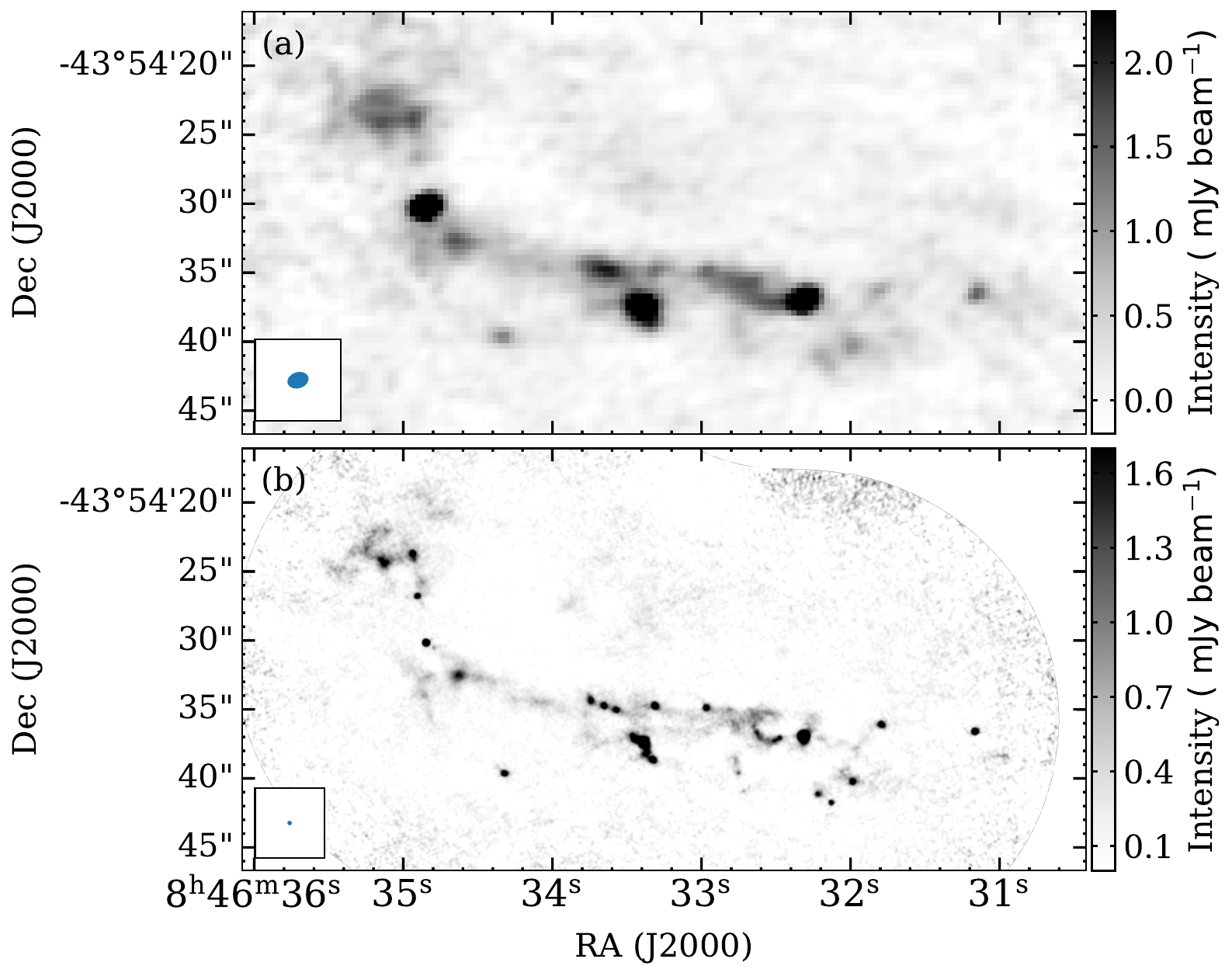} 
    \caption{ATOMS 3-mm (panel\,a) and QUARKS 1.3-mm (panel\,b) dust continuum images without any labels.  } 
    \label{fig:band3:band6}    
\end{figure*}

\begin{figure*}[ht!]
    \centering
    \includegraphics[angle=0, width=0.5\textwidth]{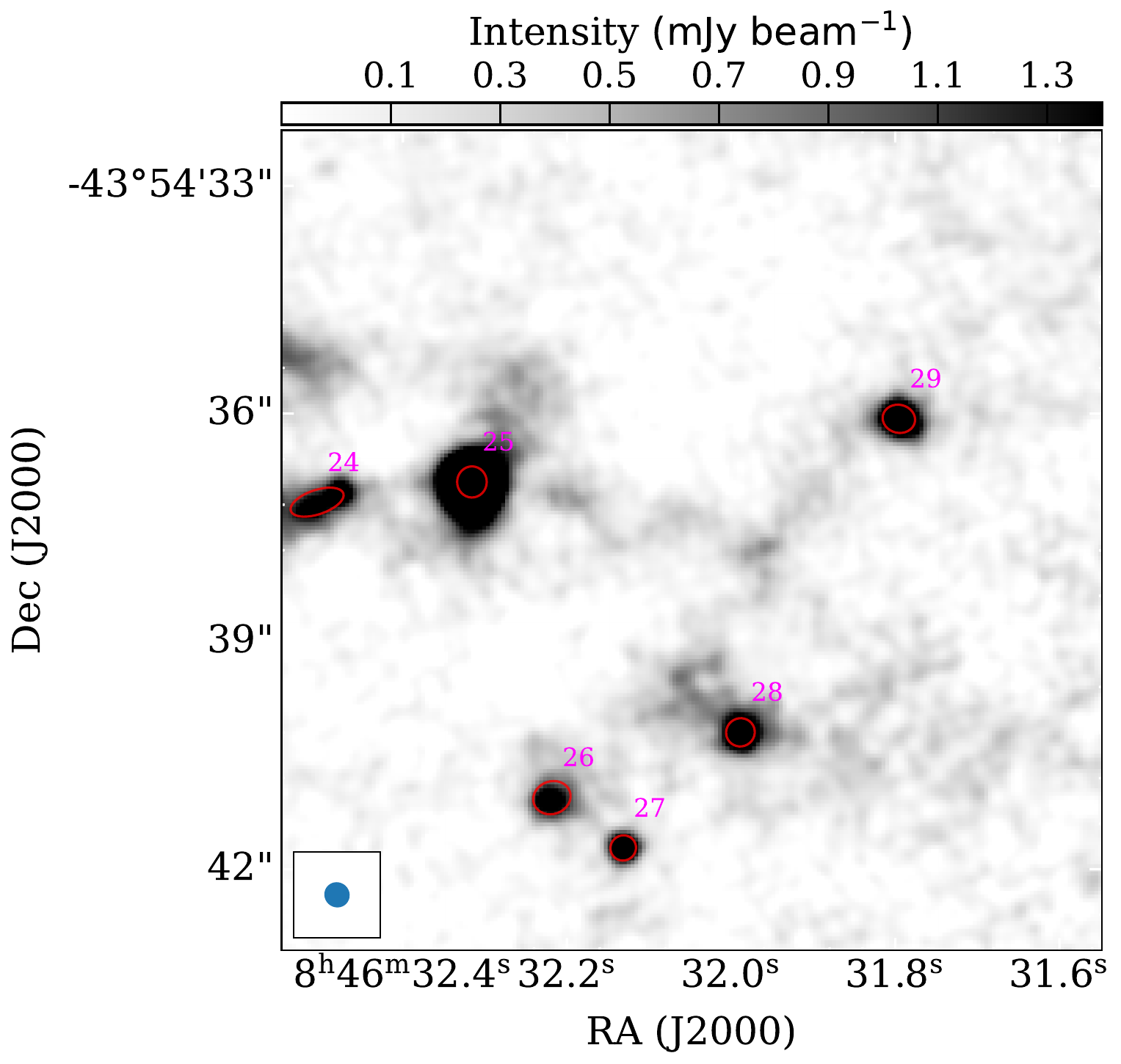} 
    \caption{Zoomed-in map of the 1.3-mm dust continuum. A network of fibers connecting condensations 25, 28, and 29 is observable but not identified automatically by {\it getsf}.} 
    \label{fig:band6:zoomin}    
\end{figure*}

\begin{figure*}[ht!]
    \centering
    \includegraphics[angle=0, width=0.8\textwidth]{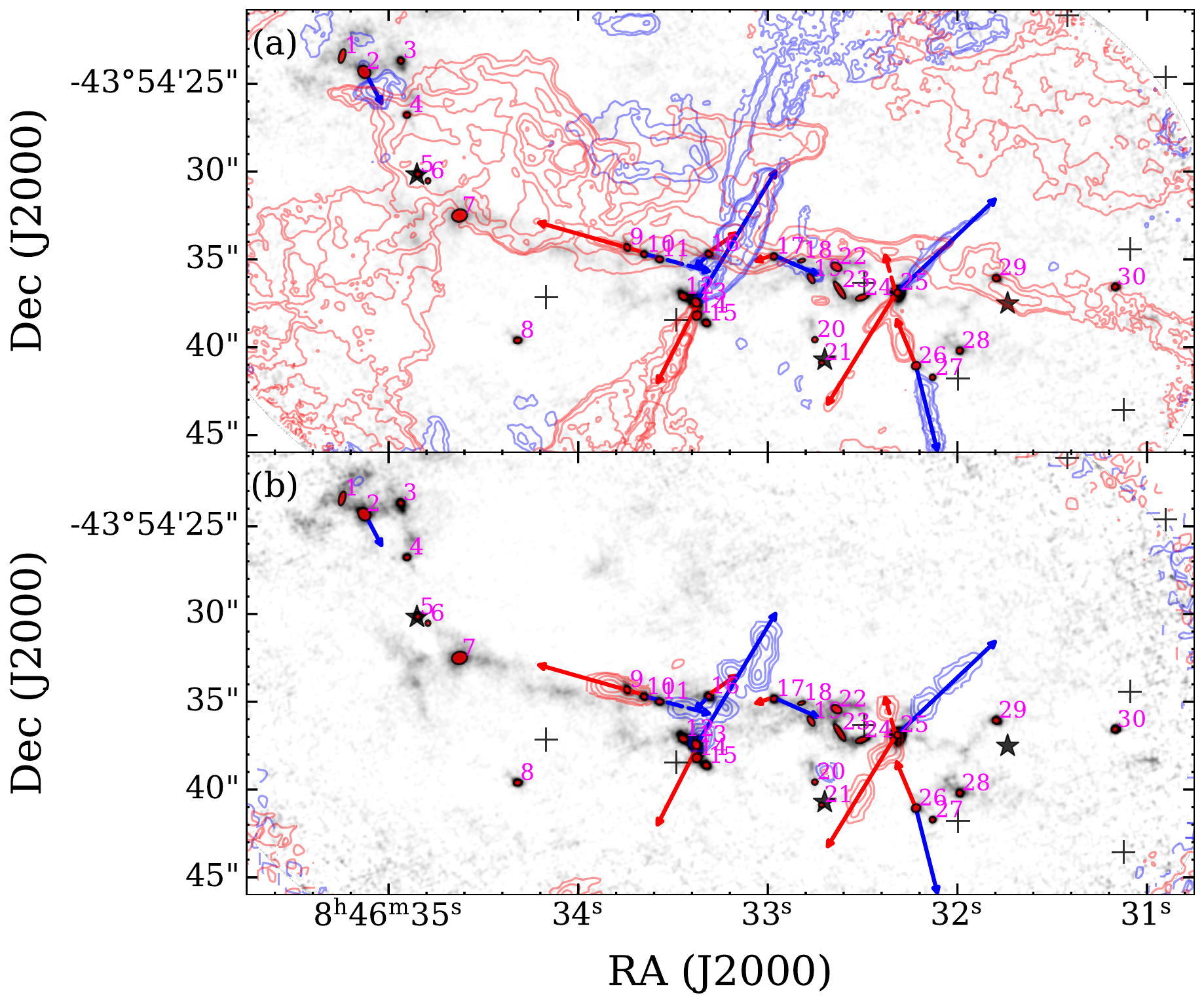} 
    \caption{CO and SiO outflows overlaid on the 1.3\,mm dust continuum. Panel\,(a): CO (2--1) outflows. The contours show [3, 6, 12, 24, 48]\,$rms$ ($\sim 0.04\,\rm Jy\,beam^{-1}\,km\,s^{-1}$). Panel\,(b): SiO (5--4) outflows. The contours same as panel\,(a), but $rms$ of $\sim \rm 0.015\,Jy\,beam^{-1}\,km\,s^{-1}$.} 
    \label{fig:outflow}    
\end{figure*}

\begin{figure*}[ht!]
    \centering
    \includegraphics[angle=0, width=0.8\textwidth]{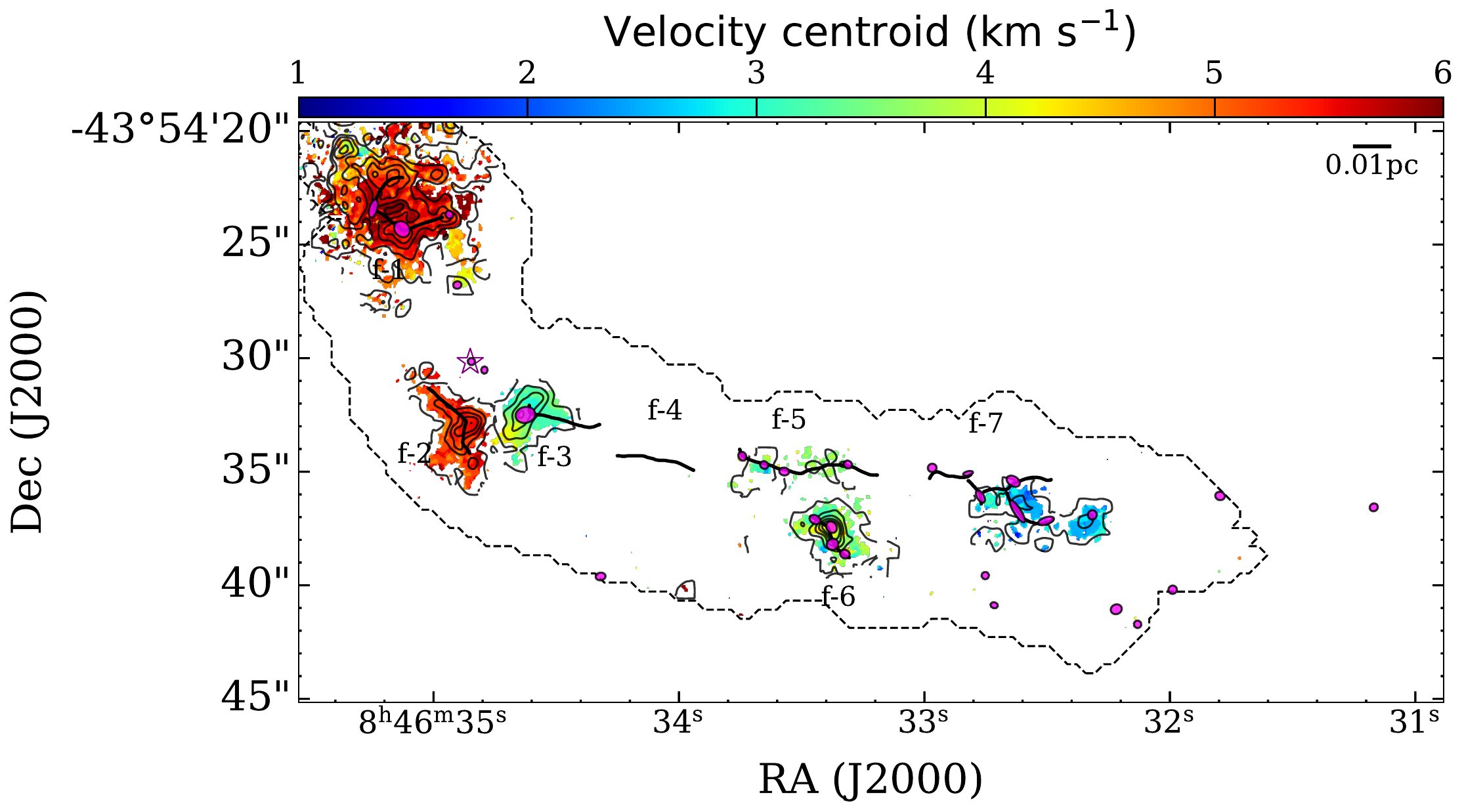} 
    \caption{Velocity centroid (Moment\,1) map of DCN line data from the QUARKS survey. The velocity range of DCN emission is set to [1, 6]\,$\rm km\,s^{-1}$. The symbol same as Fig\,\ref{fig:h13co:moment}, the contour levels start at 3 $rms$ ($\sim\rm 0.004\,Jy\,beam\,km\,s^{-1}$), and following as [6, 9, 12, 15]\,$rms$.}  
    \label{fig:dcn:moment1}    
\end{figure*}

\clearpage


\begin{deluxetable*}{lcccccccccc}
\tabletypesize{\footnotesize}
\tablecaption{Physical parameters of 1.3\,mm condensations \label{tab:core}}
\tablewidth{0pt}
\tablehead{
Condensation &  \multicolumn{2}{c}{\underline {~~~~~~~~~~Position~~~~~~~~~~}}    &  $ FWHM_{\rm maj}$  &       $ FWHM_{\rm min}$       &   PA          &     $S^{\rm int}_{\rm 1.3\,mm}$  &      $F^{\rm peak}_{\rm 1.3\,mm}$ &  $ R_{\rm c}$  &  $M_{\rm c}$ & $n_{\rm c}$  \\
           &   \colhead{$\alpha$(J2000)} & \colhead{$\delta$(J2000)}         &  $(\arcsec)$ & $(\arcsec)$ &    $(\degree)$  &     $\rm (mJy)$  &  $\rm (mJy~beam^{-1})$&  $\rm (au)$  &    (\msun)    &  $(\rm \times 10^8\,cm^{-3})$   }
\startdata
$\rm C1^*$ & 08:46:35.24 & -43:54:23.41 & 0.83 & 0.35 & 164.7 &  $3.0\pm0.3$ & $1.2\pm0.1$ &222.8 & $0.08\pm0.01$ & $2.19\pm0.35$ \\
C2 & 08:46:35.13 & -43:54:24.32 & 0.77 & 0.64 & 42.4 &  $6.2\pm0.3$ & $1.3\pm0.2$ & 353.4 &$0.16\pm0.02$ & $1.1\pm0.18$ \\
C3 & 08:46:34.94 & -43:54:23.68 & 0.36 & 0.33 & 36.0 & $4.0\pm0.2$ & $2.9\pm0.2$ &210.3 &  $0.10\pm0.01$ & $3.26\pm0.52$ \\
C4 & 08:46:34.90 & -43:54:26.78 & 0.37 & 0.33 & 102.2 &  $3.7\pm0.2$ & $2.8\pm0.2$ &213.2 & $0.09\pm0.01$ & $2.81\pm0.44$ \\
C5 & 08:46:34.85 & -43:54:30.15 & 0.33 & 0.33 & 69.8 &  $11.4\pm0.1$ & $9.4\pm0.1$ &201.3 & $0.29\pm0.04$ & $10.76\pm1.71$ \\
C6 & 08:46:34.79 & -43:54:30.52 & 0.33 & 0.29 & 175.5 &  $0.6\pm0.1$ & $0.6\pm0.1$ &188.7 & $0.02\pm0.01$ & $0.90\pm0.18$ \\
C7 & 08:46:34.63 & -43:54:32.50 & 0.88 & 0.70 & 102.9 &  $6.7\pm0.2$ & $1.1\pm0.1$ & 413.2 &$0.17\pm0.02$ & $0.73\pm0.08$ \\
 C8 & 08:46:34.32 & -43:54:39.61 & 0.44 & 0.34 & 98.9 &  $6.1\pm0.1$ & $3.2\pm0.1$ &235.9 & $0.15\pm0.02$ & $3.46\pm0.44$ \\
C9 & 08:46:33.74 & -43:54:34.32 & 0.42 & 0.34 & 30.0 &  $4.1\pm0.1$ & $2.4\pm0.1$ & 230.5 &$0.10\pm0.01$ & $2.47\pm0.39$ \\
C10 & 08:46:33.65 & -43:54:34.71 & 0.37 & 0.33 & 56.7 &  $4.5\pm0.1$ & $3.1\pm0.1$ &213.2 & $0.11\pm0.02$ & $3.44\pm0.55$ \\
C11 & 08:46:33.57 & -43:54:34.99 & 0.44 & 0.35 & 83.4 &  $3.8\pm0.2$ & $2.1\pm0.2$ & 239.4 &$0.10\pm0.1$ & $2.21\pm0.35$ \\
C12 & 08:46:33.45 & -43:54:37.11 & 0.51 & 0.38 & 62.0 &  $7.0\pm0.2$ & $3.5\pm0.2$ &268.5 & $0.18\pm0.2$ & $2.82\pm0.45$ \\
C13 & 08:46:33.38 & -43:54:37.44 & 0.54 & 0.45 & 34.3 &  $19.5\pm0.3$ & $7.7\pm0.3$ &178.8 & $0.49\pm0.07$ & $25.95\pm4.13$ \\
C14 & 08:46:33.37 & -43:54:38.20 & 0.53 & 0.49 & 119.9 &  $4.1\pm0.2$ & $1.7\pm0.2$ &195.4 & $0.10\pm0.01$ & $4.06\pm0.59$ \\
C15 & 08:46:33.32 & -43:54:38.63 & 0.43 & 0.38 & 68.0 &  $6.0\pm0.2$ & $3.4\pm0.2$ &246.6 & $0.15\pm0.2$ & $3.03\pm0.48$ \\
C16 & 08:46:33.31 & -43:54:34.67 & 0.38 & 0.34 & 60.8 &  $6.2\pm0.2$ & $4.2\pm0.2$ & 219.3 &$0.16\pm0.02$ & $4.59\pm0.73$ \\
C17 & 08:46:32.97 & -43:54:34.83 & 0.39 & 0.37 & 106.7 &  $4.6\pm0.2$ & $2.9\pm0.2$ &231.7 & $0.12\pm0.02$ & $2.92\pm0.46$ \\
$\rm C18^*$ & 08:46:32.82 & -43:54:35.07 & 0.45 & 0.19 & 103.8 &  $1.1\pm0.1$ & $1.0\pm0.1$ &178.4 & $0.03\pm0.01$ & $1.60\pm0.25$ \\
$\rm C19^*$ & 08:46:32.77 & -43:54:36.09 & 0.61 & 0.31 & 29.6 &  $2.8\pm0.3$ & $1.0\pm0.1$ &265.3 & $0.07\pm0.01$ & $1.13\pm0.18$ \\
C20 & 08:46:32.75 & -43:54:39.57 & 0.34 & 0.32 & 98.4 &  $1.1\pm0.2$ & $0.9\pm0.2$ &201.2 & $0.03\pm0.01$ & $1.12\pm0.22$ \\
C21 & 08:46:32.72 & -43:54:40.88 & 0.34 & 0.27 & 76.2 &  $0.8\pm0.1$ & $0.7\pm0.1$ &184.8 & $0.02\pm0.01$ & $0.96\pm0.15$ \\
$\rm C22^*$ & 08:46:32.64 & -43:54:35.42 & 0.64 & 0.43 & 58.7 &  $4.5\pm0.5$ & $1.1\pm0.1$ &209.6 & $0.11\pm0.02$ & $3.62\pm0.57$ \\
$\rm C23^*$ & 08:46:32.62 & -43:54:36.75 & 1.15 & 0.31 & 32.8 &  $6.4\pm0.6$ & $1.7\pm0.2$ &272.4 & $0.16\pm0.03$ & $2.40\pm0.38$ \\
$\rm C24^*$ & 08:46:32.50 & -43:54:37.17 & 0.73 & 0.32 & 108.5 &  $5.2\pm0.5$ & $2.4\pm0.2$ &294.8 & $0.13\pm0.02$ & $1.54\pm0.24$ \\
C25 & 08:46:32.32 & -43:54:36.90 & 0.41 & 0.39 & 179.0 &  $51.7\pm0.4$ & $26.2\pm0.2$ &243.9 & $1.31\pm0.18$ & $27.34\pm4.33$ \\
C26 & 08:46:32.22 & -43:54:41.06 & 0.5 & 0.43 & 109.8 &  $4.5\pm0.2$ & $1.7\pm0.1$ &282.8 & $0.11\pm0.02$ & $1.47\pm0.23$ \\
C27 & 08:46:32.13 & -43:54:41.72 & 0.34 & 0.33 & 121.0 & $3.2\pm0.1$ & $2.6\pm0.1$ & 204.3 & $0.08\pm0.01$ & $2.84\pm0.45$ \\
C28 & 08:46:31.99 & -43:54:40.20 & 0.38 & 0.37 & 119.4 &  $4.2\pm0.2$ & $2.7\pm0.2$ &228.7 & $0.11\pm0.01$ & $2.78\pm0.44$ \\
C29 & 08:46:31.80 & -43:54:36.07 & 0.43 & 0.37 & 77.5 &  $6.6\pm0.1$ & $3.4\pm0.1$ & 243.3 &$0.17\pm0.02$ & $3.57\pm0.57$ \\
C30 & 08:46:31.17 & -43:54:36.57 & 0.37 & 0.34 & 124.1 & $8.8\pm0.2$ & $6.2\pm0.1$ & 216.4 &  $0.22\pm0.03$ & $6.57\pm1.04$ \\
\enddata
\begin{flushleft}
$^*$ These condensations are identified from SExtractor.
\end{flushleft}
\end{deluxetable*}

\clearpage

\bibliographystyle{aasjournal}
\bibliography{reference.bib}{}

\end{document}